\documentclass{article}
\pdfoutput=1
\usepackage[utf8]{inputenc}
\usepackage[english]{babel}
\usepackage[square, comma, sort&compress, numbers]{natbib}
\usepackage{amsfonts}
\usepackage[fleqn]{amsmath}
\usepackage{amssymb}
\usepackage{graphicx}
\usepackage{float}
\usepackage{multirow}
\usepackage{url}
\usepackage{pgfgantt}
\usepackage{subfigure}
\usepackage{caption}
\usepackage{indentfirst}
\usepackage{booktabs}
\usepackage{colortbl}
\usepackage{multicol}
\usepackage{grffile}
\usepackage{makeidx}
\usepackage{algorithm}
\usepackage{algpseudocode}
\usepackage{cases}
\usepackage{bm}
\usepackage{authblk}
\graphicspath{./figures/}
\emergencystretch=\maxdimen
\author[a,b]{Zhengyong Jiang}
\author[d]{Jeyan Thiayagalingam \thanks{Corresponding author: t.jeyan@stfc.ac.uk}}
\author[a,b,c] {Jionglong Su \thanks{Corresponding author: jionglong.su@xjtlu.edu.cn}}
\author[e]{Jinjun Liang}
\affil[a]{Department of Mathematical Sciences, Xi'an Jiaotong-Liverpool University, Suzhou, China}
\affil[b]{Department of Electrical and Electronic Engineering, University of Liverpool based in Xi'an Jiaotong-Liverpool University, Suzhou, China}
\affil[c]{Neusoft Corporation, Shenyang, China}
\affil[d]{Scientific Computing Department, Science and Technologies Facilities Council, Harwell Campus, Oxford, UK}
\affil[e]{Newzeland}
\begin{document}
\setlength{\parindent}{2em}
\setlength\parskip{.5\baselineskip}

\pagestyle{headings}

\title{CAD: Clustering And Deep Reinforcement Learning Based Multi-Period Portfolio Management Strategy}

\maketitle

\section{Abstract}
In this paper, we present a novel trading strategy that integrates reinforcement learning methods with clustering techniques for portfolio management in multi-period trading. Specifically, we leverage the clustering method to categorize stocks into various clusters based on their financial indices. Subsequently, we utilize the algorithm Asynchronous Advantage Actor-Critic to determine the trading actions for stocks within each cluster. Finally, we employ the algorithm DDPG to generate the portfolio weight vector, which decides the amount of stocks to buy, sell, or hold according to the trading actions of different clusters. To the best of our knowledge, our approach is the first to combine clustering methods and reinforcement learning methods for portfolio management in the context of multi-period trading.

Our proposed strategy is evaluated using a series of back-tests on four datasets, comprising a of 800 stocks, obtained from the Shanghai Stock Exchange and National Association of Securities Deal Automated Quotations sources. Our results demonstrate that our approach outperforms conventional portfolio management techniques, such as the Robust Median Reversion strategy, Passive Aggressive Median Reversion Strategy, and several machine learning methods, across various metrics. In our back-test experiments, our proposed strategy yields an average return of 151\% over 360 trading periods with 800 stocks, compared to the highest return of 124\% achieved by other techniques over identical trading periods and stocks.

\noindent{Portfolio Management, Reinforcement Learning, Algorithm Asynchronous Advantage Actor-Critic, Algorithm DDPG, DBSCAN clustering method}

\section{Introduction}
Portfolio management refers to the process of allocating a specific amount of funds across a diverse range of assets in order to maximize returns under the same level of risk \citep{Cooper1999New} \citep{Davelos2001Modern}. Previous research has proposed a novel strategy called RNN-LSTM that combines deep learning and constrained optimization for multi-period trading in portfolio management \citep{Jiang2018Long}. However, this strategy has two weaknesses. Firstly, portfolio updating is based on price ratio prediction, which may not always be accurate due to unpredictable events that can affect stock prices \citep{Chen2021Mean}. Secondly, the assumption that stock prices are independent may not hold in reality, as there may be an interaction effect between stock prices in the market \citep{Markowitz1952Portfolio}, which has been considered in other strategies such as OLMAR and PAMR \citep{Li2012Online} \citep{Li2012PAMR}.

The aim of this paper is to propose a strategy that combines clustering method and model-free reinforcement learning methods to address the issue of price ratio prediction in the RNN-LSTM strategy. To the best of our knowledge, this is the first time a model-free reinforcement method and the clustering method has been combined and implemented in a multi-stock setting to output trading signals that do not have explicit mapping relationships with the reward function, such as expected value and mean square variance of the portfolio. The output of our network is a trading signal vector, with each element representing a different trading signal for each stock. The advantage of the strategy proposed in this paper is four-fold. Firstly, model-free reinforcement learning methods do not require a pre-established model of the environment \citep{Tadepalli2007Model}, making them adaptable to different markets \citep{Gu2016Q}; secondly, clustering stocks into different clusters and using different neural networks to train a trading policy for each cluster can improve the effectiveness of the agent in interacting with the environment through the reward function; thirdly, constructing the output as a trading signal vector avoids the need for accurate price ratio prediction \citep{Kaiser2019Model} and allows for more flexibility in applying different trading rules; and finally, the trading signal vector can be used to measure risk quantitatively, allowing for greater control over the risk of the portfolio.

The motivation of this research is two-fold. Firstly, the reliance of RNN-LSTM strategy on price ratio prediction may result in high volatility in performance. This motivates us to apply a model-free reinforcement learning method to avoid the problem at price predicting of the RNN-LSTM strategy. Secondly, existing reinforcement learning methods output a portfolio or trading action directly, limiting the ability of agent to measure risk quantitatively. By outputting a trading signal vector, this strategy provides greater flexibility in controlling risk and allows for the commission fee to be taken into consideration. Additionally, the clustering method is employed to ensure that each different cluster of stock can be trained by different neural networks to obtain a unique trading policy.

The novelty of this paper are as follows:
\begin{enumerate}
    \item To the best of our knowledge, we are the first to combine reinforcement learning methods and the clustering method for handling the portfolio construction in the context of multi-period trading. We establish a novel model-free reinforcement learning based strategy, which uses clustering method to divide stocks into different clusters and training neural networks by using algorithm Asynchronous Advantage Actor-Critic (A3C) for each cluster; and
    \item Algorithm Deep Deterministic Policy Gradient (DDPG) is used to determiNE the appropriate amount of capital to allocate to each cluster, enable us to manage trading volume and mitigate portfolio risk;
\end{enumerate}

The key-contributions of this research are as follows:
\begin{enumerate}
    \item To the best of our knowledge, we are the first to combine reinforcement learning methods and the clustering method for handling the portfolio construction in the context of multi-period trading. Backtesting results show that the clustering method reduces the Max Drawdown from 0.148 to 0.089.
    \item We conduct a comprehensive evaluation of our strategy and validate its effectiveness by performing backtests on four datasets containing 800 real-world stocks over two different time periods. Our strategy is compared against ten other strategies, including RNN-LSTM, Robust Median Reversion (RMR), Passive Aggressive Median Reversion (PAMR), and several machine learning methods based on various metrics such as Final Value, Max Drawdown, Sharpe Ratio, Positive Days Sortino Ratio, and Calmar Ratio for each strategy.
\end{enumerate}

The remaining sections of this paper are organized as follows. Section 2 provides a discussion of the background research on portfolio construction. Next, Section 3 presents the problem statement. Then, Section 4 introduces the methodology used to construct the portfolio. The results of the back-tests are presented in Section 5. Finally, Section 6 discusses the conclusions and future work of our research.

\section{Background}
Although existing model-based deep reinforcement learning methods have made remarkable achievements in portfolio management, the majority of them rely on price prediction \citep{Heaton2017Deep, Xiong2016Deep}, which may not be highly accurate, as demonstrated by the back results from our past research \citep{Jiang2018Long}.

Previous successful reinforcement learning attempts for model-free portfolio selection schemes include variants of reinforcement learning \citep{Moody2001, Dempster2006} and deep Q-learning methods \citep{Mnih2015}. However, these methods can only output trading action grades for a single asset. In this research, we are the first, to the best of our knowledge, to apply model-free reinforcement learning and clustering methods to multiple assets, generating trading signals for stocks in different clusters that do not have explicit mapping relationships with the reward function, such as the expected value and mean square variance of the portfolio. We apply clustering methods and model-free reinforcement learning methods in our strategy for three reasons. Firstly, trading different categories of stocks together may affect the effectiveness of agent in interacting with the environment through the reward function. Clustering stocks into different categories allows us to use different neural networks to learn a trading policy for each cluster. Secondly, model-free methods avoid the need for price prediction, which is difficult to achieve with high accuracy. The ultimate goal of portfolio management is to maximize profits, not price prediction accuracy \citep{Model2019}. Finally, compared to price movement predictions or trading action grades, trading signal vectors are more flexible and can be more easily applied based on different trading rules. This is because we can use different trading rules, designed by ourselves, to convert the trading signal vector into trading actions.

\section{Problem Statement}
\begin{figure}[htb]
	\centering
	\fbox{\includegraphics[width=12cm]{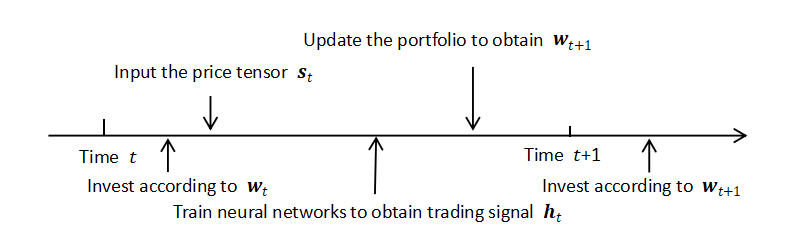}}
	\caption{The time line of our action in a single trading period.}
	\label{fig:timeline}
\end{figure}

In a financial market, suppose we are interested in investing $k$ assets continuously for a period of $n$ trading days. Our investment on $k$ assets is represented by the portfolio weight vector ${{\bf{w}}_{t}} = \left[ w_{t}^1, \, \ldots \, ,w_{t}^k \right]^T \in {\left[ {0,1} \right]^k}$, which denotes the proportion of wealth invested in asset $j \in \left\{ {1,2, \ldots ,k} \right\}$ just before ${{t}^{{\rm{th}}}}$ trading day. Here, $w_{t}^1 + w_{t}^2 +  \ldots  + w_{t}^k = 1$. After investing, we calculate the return of the investment at ${t^{{\rm{th}}}}$ trading day using a price ratio vector ${{\bf{x}}_{t}} = \left[ x_{t}^1, \, \ldots \, ,x_{t}^k \right]^T \in \mathbb{R} _ + ^k$. The element of price ratio vector is defined as $x_{t}^j = {cp}_{t}^j/{cp}_{t - 1}^j$ where ${cp}_{t}^j$ represents the close price of ${j^{{\rm{th}}}}$ asset at ${t^{{\rm{th}}}}$ trading day and the element $x_{t}^j$ represents the return of ${j^{{\rm{th}}}}$ stock at ${t^{{\rm{th}}}}$ trading day. Figure \ref{fig:timeline} illustrates the timeline of our action in a single trading period. When the trading market is closed, we collect price information such as open price, close price, highest price, and lowest price to generate a price tensor ${{\bf{s}}_{t}}$. We then use the price tensor as input to train the neural network to output the trading signal vector and obtain the portfolio weight vector ${{\bf{w}}_{t + 1}}$ for investing in the next trading day (${{t+1}^{{\rm{th}}}}$ trading day) after the portfolio updating.

In this reserach, we design a strategy for determining the portfolio vector ${{\bf{w}}_{t}}$ at the beginning of the ${t^{{\rm{th}}}}$ trading day to maximize the final value ,${V_n}$, at the end of ${n^{{\rm{th}}}}$ trading day. Here, $${V_n} = {V_0}\prod _{t = 1}^n({\bf{w}}_{t}^T{{\bf{x}}_t})$$ where ${V_0}$ is the initial value of our wealth at the beginning of trading and ${\bf{w}}_{t}^T{{\bf{x}}_{t}}$ is the period return on the $d$ assets of at the end of ${t^{{\rm{th}}}}$ trading day. This strategy is assessed based on the final cumulative portfolio wealth and other metrics, which will be introduced later in this paper.

In this study, we examine the back-testing trading scenario, where the trading agent is positioned back in time at a specific point in the market history, without any knowledge of future market information, and conducts paper trading from that point onwards. The assumptions used in our experiments are not trivial and are as follows:

\begin{enumerate}
  \item[1.] \textbf{Perfect liquidity \citep{Efficient1970}}. This assumption assumes that the market has perfect liquidity, allowing us to invest our capital in each asset with any possible proportion without considering whether there are enough order tickets in the market.
  \item[2.] \textbf{Zero impact cost \citep{Jacob2007}}. This assumption ensures that we can complete each trading immediately at the last price when the orders are put.
\end{enumerate}

These two assumptions are close to the real-world trading environment if the trading volume in a market is high enough \citep{Exit2014}.
\section{Methodology}
We present a novel approach, called the \textbf{C}lustering \textbf{A}nd Deep \textbf{R}einforcement Learning (CAD) strategy, in this paper, which combines the DBSCAN clustering method, reinforcement learning with algorithm A3C, and algorithm DDPG to address the portfolio management problem in the context of multi-period trading. Our strategy comprises four steps:

\begin{itemize}
   \item[Step 1.] \textbf{Data preprocessing step}. We extract various financial indices from daily trading records of stocks, which are used to construct the input tensor of the algorithm A3C step.
   \item[Step 2.] \textbf{Clustering step}. We partition stocks into several clusters using the T-SNE method and the DBSCAN clustering method based on the indices extracted in step 1.
   \item[Step 3.] \textbf{Algorithm A3C step}. In this step, our strategy employs the algorithm A3C to generate trading signal vectors for corresponding clusters obtained in step 2. The trading signal vector determines which stocks in the same cluster to invest in. The input tensor of the algorithm A3C is constructed in step 1. The trading signal vectors are subsequently converted into trading actions according to various predetermined trading rules.
   \item[Step 4.] \textbf{Algorithm DDPG step}. The algorithm DDPG is used to determine the number of stocks of different clusters to buy or sell by combining trading actions of various clusters obtained in step 3.
\end{itemize}

\begin{figure}[htb]
	\centering
	\fbox{\includegraphics[width=12cm]{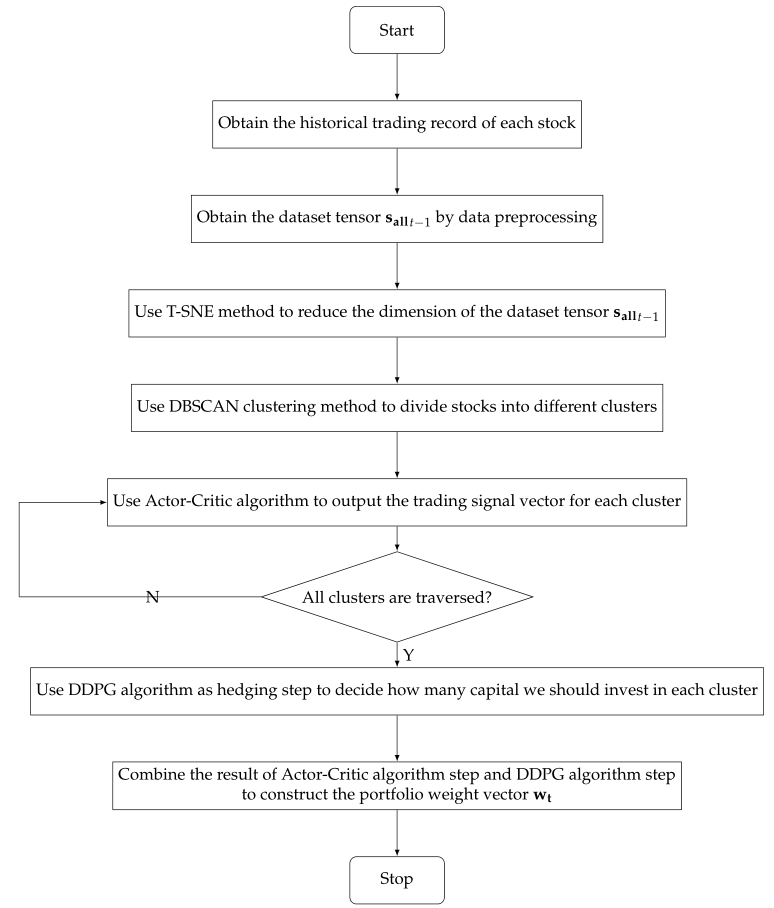}}
	\caption{The flowchart of CAD strategy.}
	\label{fig:flowchart_CAD}
\end{figure}

The flowchart of the methodology of our strategy is given in Figure \ref{fig:flowchart_CAD}. The detailed information on these steps would be given in the following text.

\subsection{Data Preprocessing}
In the Step 1 of our strategy, we preprocess the daily trading records of each stock. Specifically, we extract 25 different indices, such as the 5 Day's Moving Average price (5MA) \citep{Market2015}, Exponential Moving Average Price (EMA) \citep{Jeremy2014}, and Moving Average Convergence/Divergence (MACD) \citep{Moving2014}, from the daily trading records of each stock, including open price, close price, highest price, lowest price, and volume. These 25 indices are commonly utilized in the financial market and financial research. Subsequently, we gather these indices to construct the tensor $\bf{{s_{all}}}_{t}$ as the input of the clustering method, which contains the 25 different indices of various stocks.
\subsection{Cluster Step}
Stocks can usually be classified into different categories based on their own attributes. By adopting different trading strategies for different types of stocks, we can improve the effectiveness of our strategies. Unfortunately, due to the fact that stocks can be classified into different categories from different perspectives, we cannot accurately determine which classification method can better reflect the financial attributes of stocks. Therefore, we choose to use clustering methods to classify stocks into different categories based on financial indicators. Since the index tensor contains thousands of daily trading records of each stock, it is hard for the clustering method to deal with such a high-dimensional dataset. Hence, before the clustering step, we shall use the t-Distributed Stochastic Neighbor Embedding (T-SNE) method which is a technique for dimensionality reduction.
\subsubsection{T-SNE Method}
The T-SNE method is proposed by Laurens van der Maaten and Geoffrey Hinton in 2008 \citep{2008Visualizing} and it is used to map high-dimensional data to a low-dimensional space.
T-SNE method would try to let these two conditional probabilities to be the same so that we would map these high-dimensional data into the low-dimension space. It is also widely used in the field of the visualization of high-dimensional data \citep{2008Visualizing}. Compared with other dimension reduction methods, there are two main advantages of the T-SNE method:

\begin{itemize}
   \item[1.] \textbf{More data choice}. The T-SNE could perform well in both linear dependence and non-linear dependence high-dimensional datasets.
   \item[2.] \textbf{More Clear Distribution}. T-SNE method keeps a long distance between two high-dimensional data points with low similarity in low-dimensional space. Hence, there would be a clear boundary between data points with different categories.
\end{itemize}

We choose the T-SNE method as dimension reduction method in our research since it is the most popular dimension reduction method. Also, the T-SNE method would establish a clear boundary between data points with different categories which could help the clustering method easier to divide data points into different categories.
\subsubsection{Clustering Method}
After the dimensionality reduction is completed in the T-SNE method, the low-dimension data is used as input in the clustering method. We use the DBSCAN clustering method \citep{Cluster1996} to divide stocks into different categories.

The key idea of the DBSCAN clustering method is that for each point of a cluster, the neighbor-hood of a given radius has to contain at least a minimum number of points (MinPts), i.e., the density in the neighborhood has to exceed some threshold (eps). Algorithm \ref{alg:DBSCABN} describes the DBSCAN clustering method \citep{1996DBSCAN}.

\begin{algorithm}[H]
  \caption{DBSCAN clustering method}
  \label{alg:DBSCABN}
  \begin{algorithmic}[1]
    \Require
      Dataset $D$; Cluster radius $eps$; Minimal number of points of a cluster $MinPts$;
    \Ensure
      Several numbers of cluster $C$
    \State We denote the Eps-neighborhood of a point $p$ as ${N_{{Eps}}}\left( p \right)$. The Eps-neighborhood of a point is defined by ${N_{{Eps}}}\left( p \right) = \left\{ {q \in D|dist\left( {p,q} \right) \le Eps} \right\}$ where $dist\left( {p,q} \right)$ represents the Euclidean distance between point $p$ and $q$;
    \label{Eps-neighborhood of a point}
    \State We define that a point $p$ is directly density-reachable from a point $q$ w.r.t. Eps, MinPts if
    \begin{itemize}
       \item[1.] $p \in {N_{eps}}\left( q \right)$
       \item[2.] $\left| {{N_{Eps}}\left( q \right)} \right| \ge MinPts$
    \end{itemize}
    \label{directly density-reachable}
    \State We define that a point $p$ is density-reachable from a point $q$ w.r.t. $Eps$ and $MinPts$ if there is a chain of points ${p_1}, \ldots ,{p_n}$, ${p_1} = q$, ${p_n} = p$ such that
    ${p_{i+1}}$ is directly density-reachable from $p_i$ where $i \in \left\{ {1, \ldots ,n - 1} \right\}$.
    \label{density-reachable}
    \State We define that a point $p$ is density-connected to a point $q$ w.r.t. $Eps$ and $MinPts$ if there is a point $o$ such that both $p$ and $q$ are density-reachable from $o$ w.r.t. $Eps$ and $MinPts$
    \label{density-connected}
    \State We define that a cluster $C$ w.r.t. $Eps$ and $MinPts$ is a non-empty subset of $D$ satisfying the following conditions:
    \begin{itemize}
       \item[1.] $\forall p,q$: if $p \in C$ and $q$ is density-connected from $p$ w.r.t. $Eps$ and $MinPts$, then $q \in C$;
       \item[2.] $\forall p,q \in C$: $p$ is density-connected to $q$ w.r.t. $Eps$ and $MinPts$.
    \end{itemize}
    \label{cluster}
    \State Let ${C_1}, \ldots {C_k}$ be the clusters of the dataset $D$ w.r.t. $Eps$ and $MinPts$, then we denote the noise point as the points not belong to any cluster $C_i$, i.e., $p = \left\{ {p \in D|\forall i:p \notin {C_i}} \right\}$ where $i = 1,2, \ldots ,k$.
    \label{noise point}
    \State Traversing points in the dataset to obtain all $C_i$ which meet conditions in step 5 and noise points which meet conditions in step 6.
    \label{traversing the dataset} \\
    \Return $C_i$ and all noise points;
  \end{algorithmic}
\end{algorithm}

Compared to the traditional $k$-means clustering method, the DBSCAN clustering method has two main advantages:

\begin{itemize}
   \item[1.] \textbf{Automatic clustering}. DBSCAN clustering method can determine the number of clusters of the dataset automatically while the traditional $k$-means cluster method requires the number of clusters of the dataset to be known. It is the main advantage of the DBSCAN cluster method since it is difficult for researchers to know a prior the number of clusters the dataset \citep{1996DBSCAN}.
   \item[2.] \textbf{More data choice}. The DBSCAN clustering method can automatically identify clusters of arbitrary shapes because its basic idea is based on density. Compared to traditional K-Means clustering method, which can only identify convex-shaped clusters, DBSCAN can identify non-convex clusters. In the DBSCAN clustering method, a core point is defined as a point whose neighborhood of a certain density radius $\epsilon$ contains a minimum number of data points. By connecting the reachable relationships between core points, neighboring core points can form a cluster. In addition, boundary points can also be assigned to a cluster, making the shape of the cluster more flexible.\citep{1996DBSCAN}.
\end{itemize}

We choose the DBSCAN clustering method in our research since it could cluster a large number of data without a fixed number of categories. In our experiments, there are a large number of stocks needed to be clustered and we could not know the exact number of categories of these stocks. Hence, the DBSCAN clustering method is suitable for our experiments since it could generate the number of clusters automatically.
\subsection{Algorithm A3C step}
\begin{figure}[H]
 \centering
 \includegraphics[height=6cm]{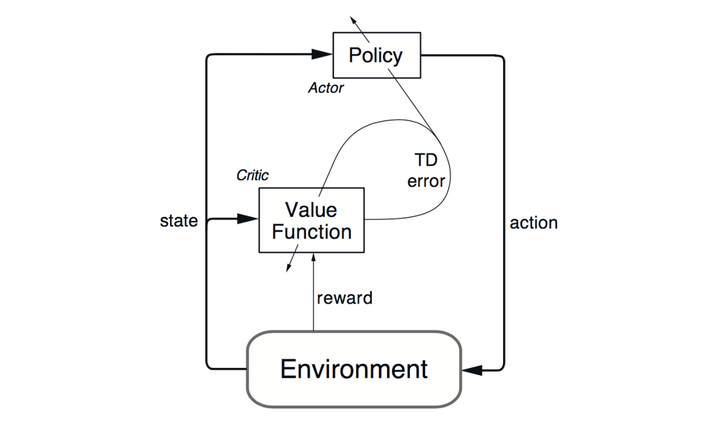}
 \caption{The structure of the Actor-Critic block.}

 \label{fig:AC}
\end{figure}

After clustering, the stocks that do not belong to any cluster will be excluded from the trading pool. The remaining stocks, which are partitioned into different clusters, would then undergo the algorithm A3C step. The algorithm A3C, which combines the advantages of the algorithm policy gradient and algorithm Q-learning, is applied to each cluster to explore an appropriate trading policy. The algorithm A3C is a variant of algorithm Actor-Critic (AC), it uses asynchronous training to simultaneously train multiple AC block to update the parameters of the neural network. The architecture of AC block is illustrated in Figure \ref{fig:AC} \citep{A2015}. This block consists of two networks, namely the Actor network and Critic network. The Actor network, which is based on the algorithm Policy Gradient \citep{Policy1999}, generates the trading signal, while the Critic network is trained to estimate the value of the portfolio and grade for the trading action. Moreover, a trading rule is utilized to convert the output of the Actor network into specific trading actions. These actions are employed to interact with the environment to obtain the reward, which is subsequently used to calculate the Temporal-Difference (TD) error with the value function. Finally, the TD error is applied to update the parameters of the Actor network and the Critic network.

Table \ref{tab:symbol} gives part of symbols which would be used in the algorithm A3C.

\begin{table}[H]
  \small
  \centering
  \setlength{\tabcolsep}{0.5mm}{
    \begin{tabular}{cc}
    \toprule
    \bf{Symbol} & \bf{Meaning of the symbol} \\
    \midrule
    $s_t$    & The state of the agent at time $t$\\
    \midrule
    $\bf{s}_t$ & The tensor we obtained in data pre-processing step \\ & which is used to represent the state at time $t$ \\
    \midrule
    $\pi \left( {a|{s_t}} \right)$     & The probability distribution of the action under the state $s_t$ \\
    \midrule
    $V\left( {{s_t}} \right)$     & The value function which is used to \\ & estimate the value of the state at time $t$ \\
    \midrule
    ${\theta _p}$ & Weight parameter used in actor network \\
    \midrule
    ${{\bf{w}}^v}$ & Weight parameter used in actor network \\
    \midrule
    $\bf{h}_t$ & The output of algorithm A3C at time $t-1$ \\
    \midrule
    $V\left( {{s_t}} \right)$ & The value function of the critic network \\
    \midrule
    $n\_c$ & The number of clusters obtained in one dataset \\
    \midrule
    $n\_c\_i$ & The number of stocks in the $i^{\rm{th}}$ cluster \\
    \midrule
    $n\_t$ & The number of trading records \\ & which we used in the training of the neural network \\
    \midrule
    $n\_f$ & The number of financial indexes \\ & which we used in the training of the neural network \\
    \bottomrule
    \end{tabular}}%
  \caption{The list of symbols which would be used in the algorithm A3C}
  \label{tab:symbol}%
\end{table}%

\subsubsection{Algorithm A3C (Actor Network)}
The input of the actor network is the tensor ${\bf{s}}_t^i$ of each cluster $i$ which is obtained from the data preprocessing step and cluster step. Specifically, the tensor constructed in step 1 contains 25 different financial indexes with an additional representing the proportion of our capital invested in each stock so that this tensor can be seen as the state of our portfolio and we could influence the state by our trading. Hence, there are 26 indices in total in the input tensor of the algorithm A3C step.

Our actor network outputs a trading signal vector ${{\bf{h}}_{t-1}} = \left[ {h_t^1,h_t^2, \ldots ,h_t^{n\_c\_i}} \right]$ at time $t-1$ for $i^{\rm{th}}$ cluster and the portfolio vector ${{\bf{b}}_{t - 1}}$ is updated according to the trading signal vector. The trading signal vector $\bf{h}_{t-1}$ is a vector whose elements consist of -1, 0, 1 which represents three types of trading signals, i.e., BUY, SELL and HOLD respectively. The size of ${{\bf{b}}_{t - 1}}$ is $n\_c\_i \times 1$ where $n\_c\_i$ represents the number of stocks in the $i^{\rm{th}}$ cluster. Trading actions would be created based on trading signals and the trading rule to trade corresponding stocks and the trading rule which would be introduced in the following section. After the trading, we would obtain a new portfolio and the reward of this trading could be calculated.

Our actor network is based on the algorithm Policy Gradient \citep{Policy1999} to generate trading signals. According to our trading rule, the agent will sell/buy stocks when the corresponding trading signal is SELL/BUY. After each stock trading, the possibility of the corresponding trading signals ($\pi \left( {a|{s_{t-1}}} \right)$) is adjusted based on the value function obtained from the critic network. The critic network, which is based on the algorithm Q-learning \citep{A20152}, estimates the value of each state and assigns a grade to the trading action to evaluate whether it will increase or decrease the value of the state ($V\left( {{s_t}} \right) > V\left( {{s_{t - 1}}} \right)$). The actor network updates its parameters ${{\bf{w}}^\theta }$ based on the estimated grade from the critic network. This update increases the possibility of trading actions that would increase the value of the state and decreases the possibility of trading actions that would decrease the value of the state.

\begin{figure}[H]
 \centering
 \includegraphics[height=6cm]{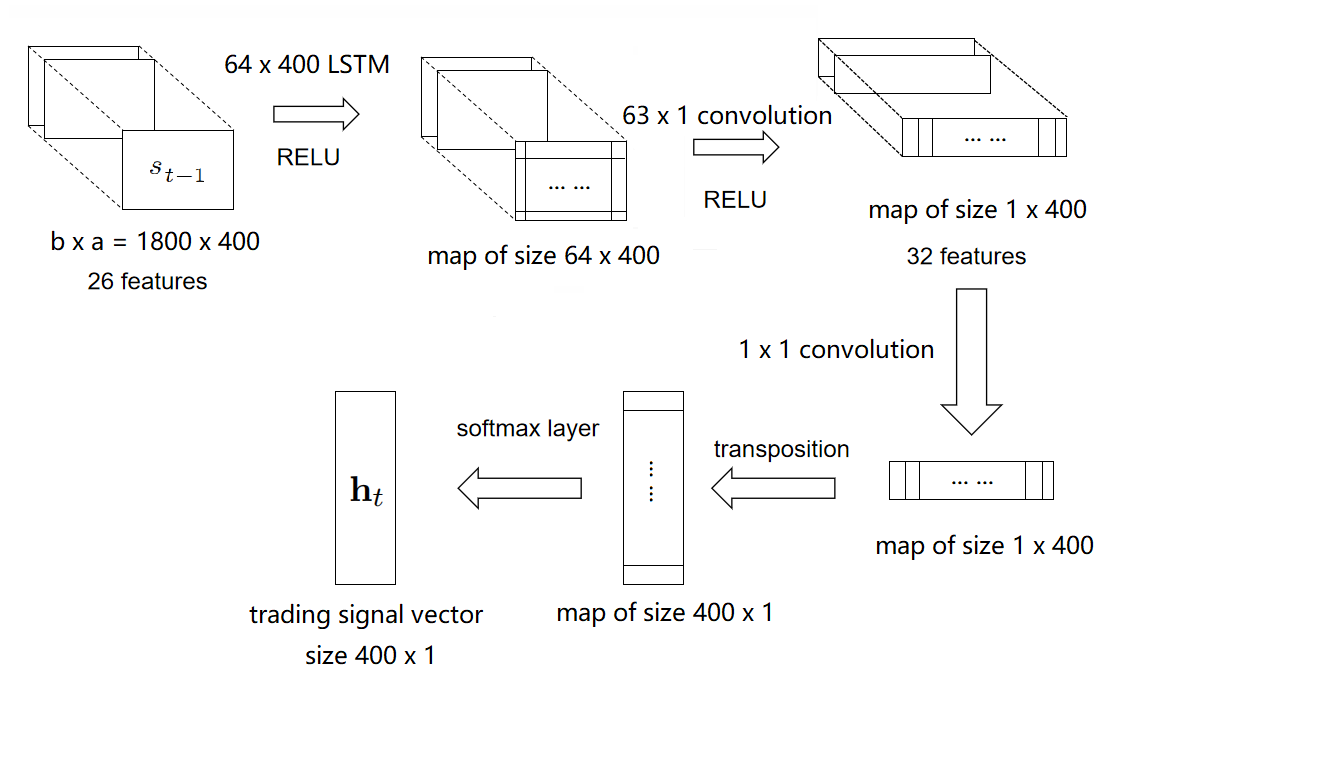}
 \caption{The architercure of the Actor network.}

 \label{fig:actor}
\end{figure}

The architecture of the actor network is depicted in Figure \ref{fig:actor}. At time $t-1$, the input of the actor network is the tensor ${{\bf{s}}_{t-1}}$, with dimensions ${n\_t} \times {n\_c\_i} \times {n\_f}$. Here, $n\_c_i$ represents the number of stocks in the $i^{\rm{th}}$ cluster, $n\_t$ represents the number of trading records used for training the actor network, and $n\_f$ represents the number of indexes utilized in the experiment. In Figure \ref{fig:actor}, we set ${n\_t}$ to be 1800, ${n\_c\_i}$ to be 400, and ${n\_f}$ to be 26. The indices used in the experiment consist of various categories, including open price, close price, highest price, lowest price, as well as several indicators such as 5 Day's Moving Average price (5MA), Exponential Moving Average Price (EMA), and Moving Average Convergence / Divergence (MACD), among others. The input tensor ${{\bf{s}}_{t-1}}$ represents the state of the agent in the reinforcement learning step. This state is fed into a recurrent neural network with LSTM block, combined with a convolution layer and a softmax layer. The resulting output is an action probability distribution $\pi \left( s_{t-1} \right)$ based on the state $s_{(t-1)}$ of the portfolio. The action with the highest probability is chosen to generate a trading signal vector ${{\bf{h}}_t} = \left[ {h_t^1,h_t^2, \ldots ,h_t^{n\_c\_i}} \right]$ at time $t-1$, and the portfolio vector ${{\bf{b}}_{t - 1}}$ is updated accordingly. The trading signal vector $\bf{h}_t$ is a ${n\_c\_i} \times 1$ vector, where each element represents one of three trading signals: BUY, SELL, or HOLD, denoted by the values -1, 0, and 1, respectively.

\subsubsection{Algorithm A3C (Critic Network)}
\begin{figure}[H]
 \centering
 \includegraphics[height=3cm]{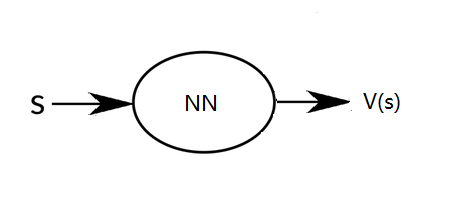}
 \caption{The structure of the critic network of algorithm A3C.}

 \label{fig:critic}
\end{figure}
The architecture of the critic network is illustrated in Figure \ref{fig:critic}. Similar to the actor network, the input of the critic network is the state tensor $\bf{s}_t$, which represents the state of the portfolio at time $t$. It is important to note the distinction between the state tensor $\bf{s}_t$ and the state $t$ of the portfolio at time $t$. The output of the critic network is the value of the state, denoted as the value function $V(s_t)$. This value function is defined as
$$V\left( {{s_t}} \right) = {E_{\pi \left( {{s_t}} \right)}}\left[ {r + \gamma  \times V\left( {{s_{t + 1}}} \right)} \right]$$.
This implies that the value of the state at time $t$, denoted as $V(s_t)$, is equal to the expected value of all possible states at time $t+1$ obtained by following the trading action probability distribution $\pi \left( s_{t} \right)$. Additionally, it includes the reward rr obtained based on the return of the portfolio ${\bf{b}}_{t}$ at time $t$ and the transaction cost incurred during the updating process from ${\bf{b}}_{t-1}$ to ${\bf{b}}_t$. The learning rate of the neural network, denoted as $\gamma$, is typically set to 0.001.
\subsubsection{Algorithm A3C Network updating}
In this section, we will discuss the objective function $J\left( \pi  \right)$ that is employed in the updating process of the actor network and the critic network in the algorithm A3C. Moreover, we will introduce the policy loss of the actor network and the loss function of the critic network.
The objective function of the actor network, denoted as $J\left( \pi  \right)$, represents the expected reward that an agent can attain under the policy $\pi$ across all possible initial states \citep{Policy1999}, i.e.,
\begin{eqnarray}
J\left( \pi  \right) = E\left[ {V\left( {{s_0}} \right)} \right].
\end{eqnarray}
which is defined as the expected return that obtained based on policy $\pi$ from the beginning state $s_0$ at time 0.
The gradient of the objective function is given by Richard Sutton \citep{Policy1999} as follows
\begin{eqnarray}
{\nabla _\theta }J\left( \pi  \right) = E\left[ {T\left( {s,a} \right) \cdot {\nabla _\theta }\log \pi \left( {a|s} \right)} \right]
\end{eqnarray}
where $T\left( {s,a} \right)$ is the Temporal-Difference (TD) error defined as \citep{Policy1999}
$$T\left( {s_t,a} \right) = r + \gamma  \times V\left( {{s_t}} \right) - V\left( {{s_{t - 1}}} \right)$$.
We would like to maximize the objective function $J\left( \pi  \right)$ and minimize the policy loss, so the policy loss ${L_\pi }$ of actor network is taken to be the negative function of $J\left( \pi  \right)$, i.e.,
\begin{eqnarray}
{L_\pi }{\rm{ = }} - J\left( \pi  \right).
\end{eqnarray}
Hence, we could maximize the objective function $J\left( \pi  \right)$ by minimizing the policy ${L_\pi }$. According to Richard Sutton \citep{Policy1999}, the objective function may be rewritten as
\begin{eqnarray}
J\left( \pi  \right){\rm{ = }}E\left[ {T\left( {s,a} \right) \cdot \log \pi \left( {a|s} \right)} \right].
\end{eqnarray}
Expanding equation (5) and substituting into equation (4), the policy loss of the actor network could be written as
\begin{eqnarray}
{L_\pi }{\rm{ =  - }}{1 \over n}\sum\limits_{i = 1}^n {T\left( {{s_i},{a_i}} \right)}  \cdot \log \pi \left( {{a_i}|{s_i}} \right)
\end{eqnarray}
Next, we shall talk about the loss function of the critic network. It is claimed \citep{Policy1999} that the truth value function $V\left( s \right)$ is satisfied the Bellman Equation
\begin{eqnarray}
V\left( {{s_0}} \right) = {r_0} + \gamma {r_1} + {\gamma ^2}{r_2} +  \ldots  + {\gamma ^{n - 1}}{r_{n - 1}} + {\gamma ^n}V\left( {{s_n}} \right).
\end{eqnarray}
The loss function $e$ of critic network can be calculated using equation (7)
\begin{eqnarray}
e = {r_0} + \gamma {r_1} + {\gamma ^2} {r_2} +  \ldots  + {\gamma ^{n - 1}} {r_{n - 1}} + {\gamma ^n} V\left( {{s_n}} \right) - V\left( {{s_0}} \right)
\end{eqnarray}
In summary, this section provides an introduction to the process of updating the actor network and critic network. Our aim is to maximize the objective function $J\left( \pi  \right)$, which represents the expected reward from the initial state. Consequently, we seek to minimize the policy loss ${L_\pi }$ in equation (5). Additionally, the critic network is utilized to estimate the value function, and thus, we aim to minimize the loss function described in equation (7) to enhance the efficacy of the critic network.

\subsubsection{The policy of portfolio updating}
In the algorithm A3C step, the trading rule discussed in section 1 is incorporated with the trading signal vector to update the portfolio, proposed in this paper for the first time, involves selling stocks indicated by a SELL trading signal and using the proceeds to purchase stocks indicated by a BUY trading signal. More information regarding the trading policy can be found in Algorithm \ref{alg:updating}.

\begin{algorithm}[H]
  \caption{The trading rule used in Portfolio Updating}
  \label{alg:updating}
  \begin{algorithmic}[1]
    \Require
      Portfolio vector $\bf{b}_{t-1}$ at time $t-1$ ; Trading sinal vector $\bf{h}_{t}$;
    \Ensure
      Portfolio vector $\bf{b}_{t}$ at time $t$
    \State Construct new vector ${\bf{h}}_{buy} = \left[ {h_{buy}^1,h_{buy}^2, \ldots ,h_{buy}^{n\_t}} \right]$, the element of new vector ${h}_{buy}^{i} = h_t^i$ when $h_t^i = 1$ and $h_{buy}^{i} = 0$ when $h_t^i \not= 1$ for $i = 1,2, \ldots ,{n\_t}$;
    \label{code:fram:Construct buy vector}
    \State Construct new vector ${\bf{h}}_{sell} = \left[ {h_{sell}^1,h_{sell}^2, \ldots ,h_{sell}^{n\_t}} \right]$, the element of new vector $h_{sell}^{i} = 1$ when $h_t^i = -1$ and $h_{sell}^{i} = 0$ when $h_t^i \not= -1$ for $i = 1,2, \ldots ,{n\_t}$;
    \label{code:fram:Construct sell vector}
    \State Calculate the capital $R$ obtained by selling stocks with corresponding SELL signal by $R = {{\bf{b}}_{t - 1}}^T{{\bf{h}}_{sell}}$;
    \label{code:Calculate sold profit}
    \State Denote the number of positive elements in ${\bf{h}}_{buy}$ as $n$, calculate the proportion we would invest in each stock with corresponding BUY signal and update the buying vector ${{\bf{h}}_{buy}}^{'} = \left[ {h{{_{buy}^1}^{'}},h{{_{buy}^2}^{'}}, \ldots ,h{{_{buy}^d}^{'}}} \right]$, the element of new vector is equal to $h_{buy}^i{'} = h_{buy}^i \times {R \over n}$ ;
    \label{code:fram:Calculate how much should buy}
    \State We would invest our capital in stocks with BUY signal and the portfolio vector ${{\bf{b}}_{t}}$ is updated by $b_t^i = b_{t - 1}^i + {ts_{buy}^i}^{'} + b_{t - 1}^i \times h_{sell}^i$ for $i = 1,2, \ldots ,k$;
    \label{code:fram:update} \\
    \Return ${{\bf{b}}_{t}}$;
  \end{algorithmic}
\end{algorithm}

\subsection{Hedging step}
In the algorithm A3C step of our methodology, we employ the algorithm A3C to construct separate portfolios for each cluster obtained in the clustering step. Subsequently, we utilize the algorithm DDPG as our hedging method to determine the allocation of assets in each portfolio. The use of hedging methods in finance is well-established and serves to reduce investment risk \citep{2012Pricing}. Traditional hedging methods rely on mathematical models such as the Mean-Variance Model \citep{Markowitz1952Portfolio}, Capital Asset Pricing Model \citep{William1964}, and Black-Scholes Model \citep{1973The}. However, the effectiveness of these mathematical hedging methods can be affected by unexpected events such as price volatility of raw materials, negative news regarding related products, and logistical issues. Consequently, in this research, we calculate the proportion of capital to be invested in these different portfolios using the algorithm DDPG during the hedging step.

Table \ref{tab:symbol2} presents a selection of symbols that will be utilized in the algorithm DDPG.

\begin{table}[H]
  \small
  \centering
  \setlength{\tabcolsep}{0.5mm}{
    \begin{tabular}{cc}
    \toprule
    \bf{Symbol} & \bf{Meaning of the symbol} \\
    \midrule
    $s_t$ & The state of the agent at time $t$ \\
    \midrule
    ${{\bf{s_{new}}}_{t}}$ & The new tensor we obtained in data pre-processing of hedging step \\ & which is used to adapt the new structure of the neural network \\
    \midrule
    ${{\theta ^\mu }}$ & Weight parameter used in actor network \\
    \midrule
    ${{\theta ^Q }}$ & Weight parameter used in actor network \\
    \midrule
    $n\_c$ & The number of clusters obtained in one dataset \\
    \midrule
    $n\_c\_i$ & The number of stocks in the $i^{\rm{th}}$ cluster \\
    \midrule
    $n\_t$ & The number of trading records \\ & which we used in the training of the neural network \\
    \midrule
    $n\_f$ & The number of financial indexes \\ & which we used in the training of the neural network \\
    \midrule
    ${{\bf{w}}^{'}}$ & The output of the algorithm DDPG \\ & which would be used to construct the portfolio weight vector \\
    \bottomrule
    \end{tabular}}%
  \caption{The list of symbols which would be used in the algorithm DDPG}
  \label{tab:symbol2}%
\end{table}%

\subsubsection{Data Preprocessing for hedging step}
\begin{figure}[H]
 \centering
 \fbox{\includegraphics[width=10cm]{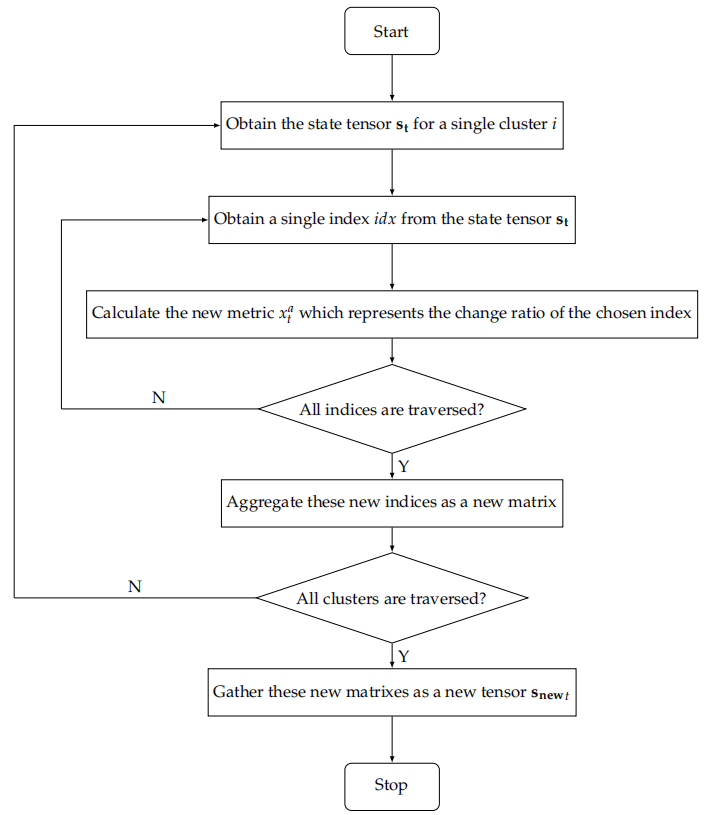}}
 \caption{The process of data preprocessing for hedging step.}

 \label{fig:flowchart_pre}
\end{figure}

The input tensor for each individual cluster $i$ in the algorithm A3C is denoted as ${{\bf{s}}_{t-1}}^i$, with a size of ${n\_t} \times {n\_c\_i} \times {n\_f}$. During the hedging step, it is necessary to preprocess each index included in the input tensor ${{\bf{s}}_{t-1}}^i$ to accommodate the new neural network and the new training target. The flowchart illustrating the data preprocessing for the hedging step is presented in Figure \ref{fig:flowchart_pre}.

We select the closing price as an illustrative example to demonstrate the calculation process of the new index ${idx}_{ratio}$, which represents the percentage change of the selected index (in this case, the closing price). This newly calculated index will be utilized to construct a tensor ${{\bf{s_{new}}}_{t}}$, which will serve as the input for the algorithm DDPG.

\begin{itemize}
   \item[1.] \textbf{Construct the matrix}. We would construct a matrix ${{\bf{s\_close}}_{t-1}}$ which record the daily close prices of stocks in cluster $i$ and the size of ${{\bf{s\_close}}_{t-1}}$ is ${n\_t} \times {n\_c\_i}$.
   \item[2.] \textbf{Calculate the price ratio}. We would calculate the price ratio of each stock in cluster $i$ at time $t$ by $rati{o^{a,t}} = {{clos{e^{a,t}}} \over {clos{e^{a,t - 1}}}}$ where $rati{o^{a,t}}$ represent the price ratio of stock $a$ at time $t$ and ${{clos{e^{a,t}}}}$ represents the close price of stock $a$ at time $t$ and we set $rati{o^{a,0}}$ equal to 1.
   \item[3.] \textbf{Gather the price ratio}. We define a new metric $close_{ratio}$ to represent the average change ratio of close price of stocks which belong to cluster $i$ at time $t$ The $close\_ratio$ is calculated as $close\_{ratio} = \sum\limits_{a = 1}^n {rati{o^{a,t}}}$ where $n$ represents the number of stocks which belong to cluster $i$.
\end{itemize}

Subsequently, the aforementioned process is repeated to calculate 25 new indices that represent the average percentage change of 25 previous indices for stocks belonging to cluster $i$. Consequently, a new matrix of size ${n\_t \times {n\_f}}$ is obtained for each cluster $i$. Finally, these matrices from each cluster are combined to form a tensor ${{\bf{s_{new}}}_{t}}$ of size ${n\_t} \times {n\_f} \times {n\_c}$, where $n\_c$ denotes the total number of clusters in the experiment. This tensor, denoted as ${{\bf{s_{new}}}_{t}}$, serves as the input for the algorithm DDPG.
\subsubsection{Algorithm DDPG}
The algorithm DDPG has gained significant popularity in the fields of time series analysis and continuous control \citep{2015Continuous}. In this study, we employ the algorithm DDPG as the hedging method due to its superior performance compared to the algorithm A3C, especially when dealing with small datasets. However, it is worth noting that the algorithm DDPG may face challenges when applied to large datasets, as its ability to explore a vast action space can become a disadvantage. Nonetheless, the inclusion of Ornstein-Uhlenbeck (OU) noise in the algorithm helps enhance the efficiency of action exploration by the neural network \citep{2001Discussion}. To address this limitation, we utilize the algorithm A3C to generate trading signals for clusters with a large number of stocks. Subsequently, algorithm DDPG is employed as a hedging method, allowing it to handle only a few integrated datasets from different clusters. In this step, the input tensor ${{\bf{s_{new}}}_{t}}$, obtained during the data preprocessing for the hedging step, is used as the input for the algorithm DDPG. The output of the algorithm DDPG is a weight vector ${{\bf{w}}^{'}} = \left[ {w{{'}_1},w{{'}_2}, \ldots ,w{{'}_{n\_c}}} \right]$ of size $1 \times  {n\_c}$, where each element represents the proportion of capital to be invested in each cluster. Algorithm DDPG is further described in Algorithm \ref{alg:DDPG}:

\begin{algorithm}[H]
\caption{DDPG algorithm}
  \label{alg:DDPG}
  \begin{algorithmic}
    \State \textbf{Randomly} initialize critic network $Q\left( {s,a|{\theta ^Q}} \right)$ and actor network $\mu \left( {s,a|{\theta ^\mu }} \right)$ with weights ${{\theta ^Q }}$ and ${{\theta ^\mu }}$;
    \State Initialize target network ${Q^{'}}$ and ${\mu ^{'}}$ with weights ${\theta ^{{Q^{'}}}} \leftarrow {\theta ^Q}$, ${\theta ^{{\mu ^{'}}}} \leftarrow {\theta ^{\mu} }$;
    \State Initialize replay buffer $R$;
    \State Initialize the number of episode $M$
    \For{$episode=1,M$}
    \State Initialize a random process $P$ for action exploration;
    \State Receive initial observation state $s_1$;
    \Loop { ${n_\theta }$-component}
    \State Select action ${a_t} = \mu \left( {{s_t}|{\theta ^\mu }} \right) + {P_t}$ according to the current policy $\mu$ and exploration noise $P_t$;
    \State Execute action $a_t$ and observe reward $r_t$ and observe new state $s_{t+1}$;
    \State Store transition $\left( {{s_t},{a_t},{r_t},{s_{t + 1}}} \right)$ in $R$;
    \State Sample a random minibatch of $N$ transitions $\left( {{s_i},{a_i},{r_i},{s_{i + 1}}} \right)$ from $R$;
    \State Set ${y_i} = r_i + \gamma  \times {Q^{'}}\left( {{s_{i + 1}},{\mu ^{'}}\left( {{s_{i + 1}}|{\theta ^{{\mu ^{'}}}}} \right)|{\theta ^{{Q^{'}}}}} \right)$;
    \State Update Critic by minimizing the loss: $L{\rm{ = }}{1 \over N}{\sum\limits_i {\left( {{y_i} - Q\left( {{s_i},{a_i}|{\theta ^Q}} \right)} \right)} ^2}$;
    \State Update the actor policy using the sampled policy gradient:
    \State $${\nabla _{{\theta ^\mu }}}J \approx {1 \over N}\sum\limits_i {{\nabla _a}Q\left( {s,a|{\theta ^Q}} \right){|_{s = {s_t},a = \mu \left( {{s_i}} \right)}}{\nabla _{{\theta ^\mu }}}\mu \left( {s|{\theta ^\mu }} \right){|_{{s_i}}}} $$;
    \State Update the target networks:
    \State $${\theta ^{{Q^{'}}}} \leftarrow \tau  \times {\theta ^Q} + \left( {1 - \tau } \right){\theta ^{{Q^{'}}}}$$
    \State $${\theta ^{{\mu ^{'}}}} \leftarrow \tau  \times {\theta ^\mu } + \left( {1 - \tau } \right){\theta ^{{\mu ^{'}}}}$$
    \EndLoop
    \EndFor
    \label{code:fram:update} \\
  \end{algorithmic}
\end{algorithm}

\subsubsection{Construct the final portfolio}
We denote the output of the algorithm DDPG as a weight vector ${{\bf{w}}^{'}} = \left[ {w{{'}_1},w{{'}_2}, \ldots ,w{{'}_{n\_c}}} \right]$ of size $1 \times  {n\_c}$. Each element ${w{{'}_i}}$ represents the proportion of capital to be invested in the ${i^{{\rm{th}}}}$ cluster. To construct the portfolio weight vector used in trading, we combine the results of the algorithm A3C and the algorithm DDPG.

\begin{algorithm}[H]
  \caption{Portfolio Construction}
  \label{alg:Construction}
  \begin{algorithmic}
    \Require
      Portfolio vector ${{\bf{b}}^{i}}$ for the ${i^{{\rm{th}}}}$ cluster where $i \in \left[ {1,n\_c} \right]$; Weight vector ${{\bf{w}}^{'}}$;
    \Ensure
      Final Portfolio vector $\bf{W}$ at time $t$;
    \For{$i=1,n\_c$}
    \State Calcluate the portfolio vector ${\bf{w}}^i$ which elements $w{^{i\_j}}$ represents invest proportion for ${j^{{\rm{th}}}}$ stock in cluster $i$ by $w{^{i\_j}} = {w^i} \times b{^{i\_j}}$;
    \EndFor
    \State Combine all portfolio vectors ${\bf{W}}$ to obtain the final portfolio vector  ${\bf{W}} = \left[ {{{\bf{w}}^1},{{\bf{w}}^{2, \ldots ,}}{{\bf{w}}^{n\_c}}} \right]$;
    \Return $\bf{W}$;
  \end{algorithmic}
\end{algorithm}

The portfolio weight vector for each time $t$ is constructed according to Algorithm \ref{alg:Construction}, and capital is invested based on this portfolio. In the next section, we introduce the process of hyper-parameter tuning for both the algorithm A3C and the algorithm DDPG, and present the results obtained.

\subsection{Hyper-parameter tuning}
In our research, we utilize the algorithm A3C and the algorithm DDPG to derive the portfolio weight vector. Both the algorithm A3C and the algorithm DDPG involve various hyper-parameters, such as the length of the window, the number of hidden units in each hidden layer, and the learning rate. A detailed description of these hyper-parameters can be found in Table \ref{tab:ac} and Table \ref{tab:ddpg}. To determine the optimal values for these hyper-parameters, we conduct hyper-parameter tuning using a dataset consisting of 100 stocks from SSE and NASDAQ. We explore different combinations of hyper-parameters to assess whether the trading action vector or portfolio weight vector obtained from the neural network can enhance the value of our portfolio.

\subsubsection{Hyper-parameter tuning of algorithm A3C}
For the algorithm A3C, the output of the neural network is a trading signal vector, which represents qualitative analysis. To determine the most accurate trading action vector, we compare the accuracy of different combinations of hyper-parameters. The accuracy of the trading vector is defined as:
\begin{eqnarray}
{\alpha_{accurcay}} = {{{n_{correct}}} \over {{n_{total}}}}
\end{eqnarray}

Here, $n_{correct}$ represents the number of trading actions that increase the value of the portfolio, and $n_{total}$ represents the total number of trading actions. To assess whether a trading action would increase the value of the portfolio, we introduce a new index called the price ratio. The price ratio is defined as: $${pr}_t^j = {cp}_t^j/{cp}_{t - 1}^j,$$
If the price ratio is greater than 1, it indicates that the stock price is expected to increase in the next trading period. Conversely, if the price ratio is less than 1, the stock price is expected to decrease. Therefore, buying stocks with a price ratio larger than 1.00005 (considering the 0.05\% commission fee), selling stocks with a price ratio lower than 1.00005 (considering the 0.05\% commission fee), or holding stocks with a price ratio larger than 1 will increase the value of the portfolio. Equation (9) is used in hyper-parameter tuning to evaluate the effectiveness of trading actions obtained from our neural network based on accuracy. The accuracy increases when the number of trading actions that increase the value of portfolio increases. Therefore, a higher accuracy indicates better results from the output of the algorithm A3C. This is why accuracy is chosen as the evaluation metric for hyper-parameter tuning to test the effectiveness of the algorithm A3C.

\begin{table}[htb]
  \centering
    \begin{tabular}{|l||r|}
      \hline
      {\bf Length of window} &                        64\\
      \hline
      {\bf Number of hidden layers} & 2 \\
      \hline
      {\bf Hidden unit of first layer} & 32 \\
      \hline
      {\bf Hidden unit of second layer} & 64 \\
      \hline
      {\bf Learning rate} & 0.0001 \\
      \hline
      {\bf Batch Size} & 16 \\
      \hline
      {\bf Type of regularization} & L1 regularization \\
      \hline
      {\bf Regularization Parameter} & 0.001 \\
      \hline
      {\bf Number of iterations} & {50} \\
      \hline
      {\bf Optimizer} &  ADAM \\
      \hline
      {\bf Period of Training Data} &                                        2008/8/6 - 2014/1/9\\
      \hline
      {\bf Period of cross validation Data} &                                        2014/1/10 - 2014/8/28\\
      \hline
      {\bf Period of Back-test Data} &                                          2014/8/29 - 2016/3/9\\
      \hline
    \end{tabular}%
    \caption{The parameters of algorithm A3C used in the experiment. These hyper-parameters are the same for each training neural network of 800 stocks in total.}
    \label{tab:ac}%
\end{table}%

Table \ref{tab:ac} presents the relevant information regarding the use of the algorithm A3C in the reinforcement learning method during training after parameter tuning. To prevent overfitting, we incorporated L1 regularization \citep{L12007} into both the actor loss function (${L_{\pi} } + \lambda \sum\limits_{\phi = 1}^m {\left| {{\omega _{\phi}}} \right|}$) and the critic loss function ($e + \lambda \sum\limits_{\phi = 1}^m {\left| {{\omega _{\phi}}} \right|}$). In our previous discussion in Section 5.4.3, ${L_{\pi} }$ represents the actor loss function, $e$ represents the critic loss function, and $\sum\limits_{\phi = 1}^m {\left| {{\omega _\phi}} \right|} $ denotes the sum of the absolute values of $m$ weighting parameters in the hidden layer of neural network. The regularization parameter is denoted as $\lambda$.

\subsubsection{Hyper-parameter tuning of algorithm DDPG}
For the algorithm DDPG, as the output of neural network is a portfolio weight vector, which involves quantitative analysis, we compare the Sharpe Ratio \citep{Sharpe1994} of trading results based on different portfolio weight vectors. We select the combination of hyper-parameters that yields the highest Sharpe Ratio. The Sharpe Ratio $s_a$ is defined as follows:
$${s_a} = \frac{{\mathbb{E}\left[ {{R_a} - {R_f}} \right]}}{{{\sigma _a}}} = \frac{{\mathbb{E}\left[ {{R_a} - {R_f}} \right]}}{{\sqrt {\text{Var}\left[ {{R_a} - {R_f}} \right]} }}$$
Here, $R_a$ represents the asset return, $R_f$ denotes the risk-free return, ${\mathbb{E}\left[ {{R_a} - {R_f}} \right]}$ represents the expected value of the excess of the asset return over the benchmark return, and ${{\sqrt {\text{Var}\left[ {{R_a} - {R_f}} \right]} }}$ is the standard deviation of the asset excess return. In finance, the Sharpe ratio (also known as the Sharpe index, the Sharpe measure, and the reward-to-variability ratio) is used to evaluate the performance of an investment by considering its risk \citep{Sharpe1994}. The ratio measures the excess return (or risk premium) per unit of deviation in an investment asset or trading strategy, which is commonly referred to as risk. The Sharpe ratio assesses how well the return of an asset compensates the investor for the risk taken. Therefore, a higher Sharpe Ratio indicates better performance of the portfolio weight vector. This is the rationale behind using the Sharpe Ratio for hyper-parameter tuning to evaluate the effectiveness of the algorithm DDPG.

\begin{table}[htb]
  \centering
    \begin{tabular}{|l||r|}
      \hline
      {\bf Length of window} &                        64\\
      \hline
      {\bf Explore Noise} & OU noise \\
      \hline
      {\bf $\Gamma$} & 0.99 \\
      \hline
      {\bf $\tau$} & 0.02 \\
      \hline
      {\bf Learning rate} & 0.0001 \\
      \hline
      {\bf Batch Size} & 32 \\
      \hline
      {\bf Replay buffer} & 10000 \\
      \hline
      {\bf Max Step} & 100000 \\
      \hline
      {\bf Optimizer} &  ADAM \\
      \hline
      {\bf Period of Training Data} &                                        2008/8/6 - 2014/1/9\\
      \hline
      {\bf Period of cross validation Data} &                                        2014/1/10 - 2014/8/28\\
      \hline
      {\bf Period of Back-test Data} &                                          2014/8/29 - 2016/3/9\\
      \hline
    \end{tabular}%
    \caption{The parameters of algorithm DDPG used in the experiment. These hyper-parameters are the same for each training neural network of 800 stocks in total.}
    \label{tab:ddpg}%
\end{table}%

Table \ref{tab:ddpg} presents relevant information regarding the algorithm DDPG, which is utilized in the training process after parameter tuning. The replay buffer refers to the memory size used to store samples based on prioritized experience replay. The usage of $\gamma$ and $\tau$ has been introduced in Algorithm 3.

When tuning these hyper-parameters in Table \ref{tab:ac} and Table \ref{tab:ddpg}, we employ a state tensor of each cluster as the input for the algorithm A3C to generate a trading signal. The size of the state tensor for each cluster is ${n\_t} \times {n\_c\_i} \times {n\_f}$ where $n\_c\_i$, where $n\_c\_i$ represents the number of stocks in the $i^{\rm{th}}$ cluster, $n\_t$ represents the number of trading records used in the training process (which is equal to 1620), and $n\_f$ represents the number of indexes (which is equal to 26). Subsequently, we construct a new state tensor for each cluster as the input for the algorithm DDPG to obtain the final portfolio weight vector. The size of the state tensor for the algorithm DDPG is ${n\_t} \times {n\_f} \times {n\_c}$, where $n\_c$ denotes the number of clusters in the dataset, $n\_t$ represents the number of trading records used in the training process (which is equal to 1620), and $n\_f$ represents the number of indexes (which is equal to 25).
\section{Results and discussion}
In our experiments, a total of 800 stocks from the Shanghai Stock Exchange (SSE) or National Association of Securities Dealers Automated Quotations (NASDAQ) were utilized. These trading records can be freely downloaded from Yahoo Finance \citep{Cazzoli2016}. The detailed information on these stocks is provided in the appendix. The data is divided into three parts based on the time sequence. The first part is the training set, which is employed for training our neural networks to obtain the portfolio weight vector. The second part is the cross-validation set, utilized for tuning the hyper-parameters of the neural network. The third part is the test set, used in the back-test experiment. For each stock, there are 1620 trading records used for neural network training, 180 trading records for hyper-parameter tuning, and 360 trading records for the back-test.
\subsection{Results of Back-tests}
This section introduces six metrics that are employed to evaluate the performance of each strategy. Subsequently, the results and insights obtained from the back-test are presented.
\subsubsection{Performance Measures}
In our experiments, six metrics were employed to compare the effectiveness of various clustering methods. These metrics are

\begin{itemize}
   \item[1.] \textbf{Final Cumulative Portfolio Wealth \citep{Li20125}}. The final cumulative portfolio wealth represents the value of the portfolio at the end of the trading period, indicating the overall profit or loss. A higher final value indicates a greater return achieved by the strategy.
   \item[2.] \textbf{Positive Days \citep{Funt2010}}. Positive days refers to the percentage of trading periods that yield a positive return ($\frac{{{p_{t + 1}}}}{{{p_t}}} > 1$). A higher positive days value indicates a greater number of trading periods with positive returns obtained by the strategy.
   \item[3.] \textbf{Max Drawdown  \citep{Li2016}}. The drawdown is a measure of the decline from a historical peak in a specific variable, typically the cumulative profit or total open equity of a financial trading strategy. For instance, let $X = \left( {X\left( t \right),t \geqslant 0} \right)$ be a random process with $X\left( 0 \right) = 0$. The drawdown at time $T$, denoted as $D\left( T \right)$, is defined as:
       $$D\left( T \right){\text{ = max}}\left\{ {0,\mathop {\max }\limits_{t \in \left( {0,T} \right)} X\left( t \right) - X\left( T \right)} \right\}$$
       The maximum drawdown (MDD) up to time TT is the maximum drawdown observed over the history of the variable. The formula is:
       $$M\left( T \right) = \mathop {\max }\limits_{\tau  \in \left( {0,T} \right)} \left[ {\mathop {\max }\limits_{t \in \left( {0,\tau } \right)} X\left( t \right) - X\left( \tau  \right)} \right]$$
       This can be interpreted as the proportion of money that would be lost in the worst-case scenario during the trading period. In other words, the lower the maximum drawdown, the less money the strategy would lose during the trading period.
   \item[4.] \textbf{Sharpe Ratio \citep{Sharpe1994} \citep{Memmel2003}}. In finance, the Sharpe ratio (also known as the Sharpe index, the Sharpe measure, and the reward-to-variability ratio) is a measure used to assess the performance of an investment by taking into account its level of risk. The ratio quantifies the excess return, or risk premium, per unit of deviation in an investment asset or trading strategy, which is commonly referred to as risk. The Sharpe ratio is mathematically defined as:
       $${s_a} = \frac{{\mathbb{E}\left[ {{R_a} - {R_f}} \right]}}{{{\sigma _a}}} = \frac{{\mathbb{E}\left[ {{R_a} - {R_f}} \right]}}{{\sqrt {\text{Var}\left[ {{R_a} - {R_f}} \right]} }}$$
       In this formula, $R_a$ represents the return of the asset, $R_f$ represents the risk-free return, ${\mathbb{E}\left[ {{R_a} - {R_f}} \right]}$ represents the expected value of the excess of the asset return over the benchmark return, and ${{\sqrt {\text{Var}\left[ {{R_a} - {R_f}} \right]} }}$ represents the standard deviation of excess return of the asset.
       The Sharpe ratio provides insight into how effectively return of an asset compensates the investor for the level of risk undertaken. Therefore, a higher Sharpe ratio indicates a better risk-adjusted performance of the investment strategy.
   \item[5.] \textbf{Sortino Ratio}. In finance, the Sortino Ratio is a metric used to measure the relative performance of an investment portfolio. Similar to the Sharpe Ratio, the Sortino Ratio takes into account the volatility of returns, but it specifically focuses on downside volatility rather than overall volatility. This allows for a distinction between bad and good volatility. The Sortino Ratio is named after Sortino and Price (1994). Mathematically, the Sortino Ratio is defined as:
       $${s_a} = \frac{{E\left[ {{R_a} - {R_f}} \right]}}{{\sqrt {Var\left[ min \left({{R_a} - {R_f}},0 \right) \right]} }}$$
       In this formula, $R_a$ represents the asset return, $R_f$ represents the risk-free return, and ${E\left[ {{R_a} - {R_f}} \right]}$ represents the expected value of the excess of the asset return over the benchmark return. The denominator, ${{\sqrt {Var\left[ min \left({{R_a} - {R_f}},0 \right) \right]} }}$, represents the downside standard deviation.
       The Sortino Ratio provides insight into how effectively return of an asset compensates the investor for the level of downside risk undertaken. A higher Sortino Ratio indicates a better risk-adjusted performance of the investment strategy.
   \item[6.] \textbf{Calmar Ratio}. The Calmar Ratio is a metric used to measure the performance of an investment portfolio. It is a modified version of the Sharpe Ratio that incorporates the Max Drawdown as a measure of risk. The Calmar Ratio was developed and introduced by Terry W. Young in 1991. Mathematically, the Calmar Ratio is defined as
       $${c_a} = \frac{{E\left[ {{R_a} - {R_f}} \right]}}{{MDD_a}}$$
       In this formula, $R_a$ represents the asset return, $R_f$ represents the risk-free return, and $MDD_a$ represents the Max Drawdown during the trading period.
       The Calmar Ratio provides insight into the risk-adjusted performance of an investment strategy. A higher Calmar Ratio indicates better performance in relation to the level of risk taken during the trading period.

\end{itemize}

The Sortino Ratio and Calmar Ratio are both modified versions of the Sharpe Ratio, but they have different emphases and applications compared to the Sharpe Ratio. The Sharpe Ratio uses variance as a measure of asset risk, while the Sortino Ratio focuses solely on downside deviation. This means that the Sortino Ratio is more sensitive to the value loss of the asset, making it particularly relevant for risk-averse investors. In contrast to the Sharpe Ratio and Sortino Ratio, the Calmar Ratio utilizes the Max Drawdown as a measure of asset risk. Unlike the Sharpe Ratio or Sortino Ratio, the Calmar Ratio is not easily influenced by recent fluctuations in asset value. As a result, it is more suitable for long-term investors.
In summary, while the Sharpe Ratio, Sortino Ratio, and Calmar Ratio are all variations of risk-adjusted performance measures, they differ in their focus and applicability. The Sortino Ratio is more sensitive to downside deviation and is relevant for risk-averse investors, while the Calmar Ratio is less influenced by recent fluctuations and is more suitable for long-term investors.

\subsubsection{Results of Back-tests and Discussion}
The initial value of the back-test experiment is set to be $10^6$ specific initial value chosen, our results would remain unaffected. The commission fee is set at 0.05\% in accordance with the trading rules of the stock market \citep{Predicting2018}.

\begin{table}[H]
  \centering
  \small
  \centering
  \setlength{\tabcolsep}{0.5mm}{
    \begin{tabular}{lcccccc}
          & \multicolumn{1}{l}{Final value} & \multicolumn{1}{l}{Max Drawdown} & \multicolumn{1}{l}{Sharpe Ratio} & \multicolumn{1}{l}{Positive Days} & \multicolumn{1}{l}{Sortino Rato} & \multicolumn{1}{l}{Calmar Ratio}\\
    \toprule
    CAPM &  141.11\% & 0.093 & 1.64 & 0.532 \\
    \midrule
    LSTM-DQN & 141.73\% & 0.095 & 1.45 & 0.531  & 2.5 & 4.03 \\
    \midrule
    A3C & 148.29\% & 0.115 & 1.05 & 0.523  & 1.76 & 3.18\\
    \midrule
    DDPG & \textbf{151.29\%} & \textbf{0.089} & \textbf{1.70} & \textbf{0.542}  & \textbf{2.94} & \textbf{5.23}\\
    \midrule
    \end{tabular}}%
  \caption{Average backtest result of Mathematical hedging method (Capital Asset Pricing Model), LSTM-DQN hedging method, A3C hedging method and DDPG hedging method based on four different time periods in four datasets which contains 800 stocks in total. The performance metrics are Final Portfolio Value, Max Drawdown, Sharpe Ratio.}
  \label{tab:dqn200}%
\end{table}

Table \ref{tab:dqn200} presents the back-test results of our strategy, utilizing different algorithms as hedging methods. The metrics Final Value, Max Drawdown, Positive Days, and Sharpe Ratio are considered in the back-test, with a commission fee of 0.05\%. In addition to the DDPG method, we have also explored three alternative methods as hedging strategies. The mathematical hedging method is based on the Capital Asset Pricing Model \citep{William1964}, the LSTM-DQN method is proposed by Yuan Gao and Ziming Gao \citep{2020Application}, and the algorithm A3C has been introduced in the previous test. The Final Cumulative Portfolio Wealth metric in the table represents the ratio of the portfolio value in the last time step to the initial portfolio value. The table demonstrates that the algorithm DDPG outperforms other methods in all six metrics. Thus, it is reasonable to employ the algorithm DDPG as the hedging method for our strategy.

Following the hedging step of the CAD strategy, we conduct a back-test experiment using 800 stocks and evaluate the performance of the CAD strategy against several well-known or recently published strategies based on various metrics. Most of the strategies compared in this study were surveyed by Li and Hoi \citep{Li20125}, including the Online Moving Average Reversion Strategy (OLMAR) \citep{Li2012On}, Passive Aggressive Median Reversion Strategy (PAMR) \citep{Li2012PAMR}, Online Newton Selection (ONS) \citep{Agarwal2006}, Exponentiated Gradient (EG), and Anticor \citep{Borodin2011}, except for the Robust Median Reversion Strategy (RMR) \citep{Huang2013Robust}. For the CAD strategy, trading records from 1620 trading periods are utilized as training data, while trading records from 360 trading periods are used in the back-test experiment after the training. The input of the neural network includes open price, close price, highest price, lowest price, and various indexes such as 5 Day's Moving Average price (5MA), Exponential Moving Average Price (EMA), and Moving Average Convergence / Divergence (MACD) of all stocks. Table \ref{tab:cl200} presents the performance of the 800 stocks in terms of the metrics Final Value, Max Drawdown, Positive Days, and Sharpe Ratio in the back-test with a 0.05\% commission fee.

\begin{figure}[H]
 \centering
 \fbox{\includegraphics[width=10cm]{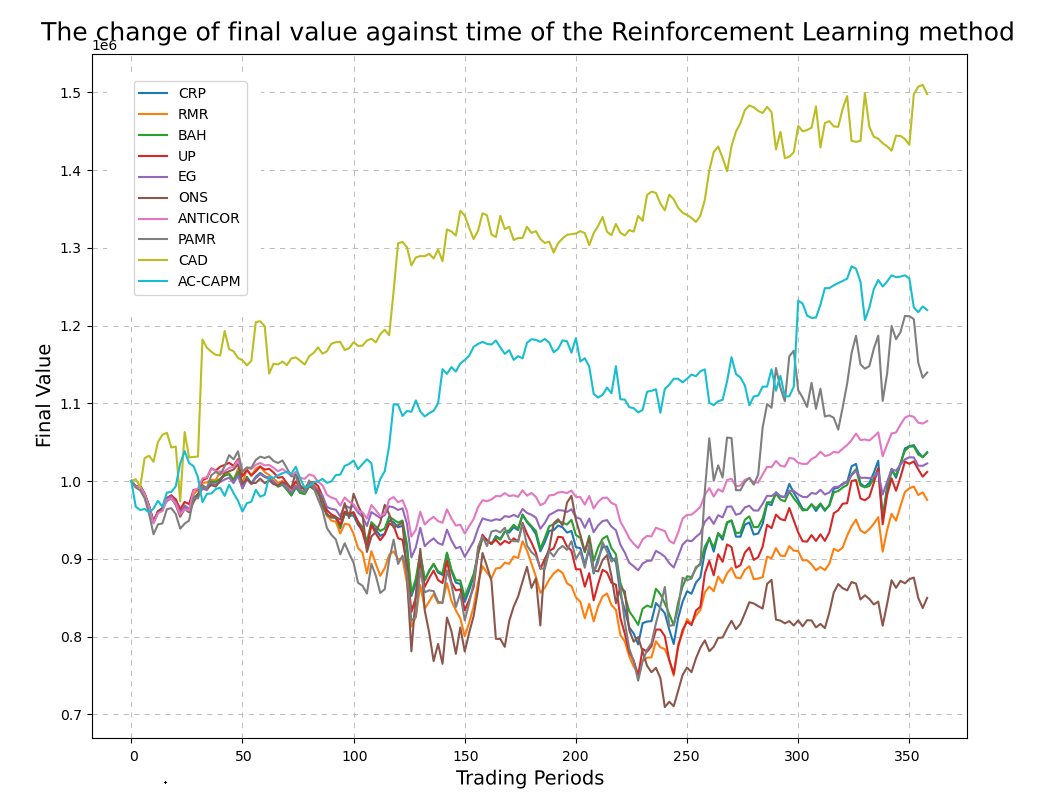}}
 \caption{The back-test result of each strategy based on one time period of one dataset which contains 200 sticks.}

 \label{fig:result_one}
\end{figure}

Fig \ref{fig:result_one} shows the plot of the back-test result based on one time period of one dataset which contains 200 sticks. We could see that in this experiment, the A3C-CAPM strategy outperforms other strategies in most of the trading periods and the CAD strategy outperforms than A3C-CAPM strategy in most of the trading periods. The satisfactory result of CAD strategy means that the CAD strategy which is based on the structure of the A3C-CAPM strategy is efficient. Next, a table would be given to show the back-test result of CAD strategy and other strategies based on all four time periods of all 4 datasets which contain 800 stocks in total.

\begin{table}[H]
  \centering
  \small
  \centering
  \setlength{\tabcolsep}{0.5mm}{
    \begin{tabular}{lcccccc}
          & \multicolumn{1}{l}{Final value} & \multicolumn{1}{l}{Max Drawdown} & \multicolumn{1}{l}{Sharpe Ratio} & \multicolumn{1}{l}{Positive Days} & \multicolumn{1}{l}{Sortino Rato} & \multicolumn{1}{l}{Calmar Ratio}\\
    \toprule
    CAD & \textbf{151.29\%} & \textbf{0.089} & \textbf{1.70} & \textbf{0.542} & \textbf{2.94} & \textbf{5.23}\\
    \midrule
    A3C-CAPM & \underline{124.06\%} & \underline{0.0929} & 1.015 & \underline{0.523} & 1.04 & \underline{2.58}\\
    \midrule
    ANTICOR & 107.77\% & 0.175 & 0.705 & 0.503 & 1.01 & 0.499 \\
    \midrule
    BAH  & 101.54\% & 0.171 & 0.289 & 0.501 & 0.32 & 0.148 \\
    \midrule
    CRP & 109.22\% & 0.148 & 0.833 & 0.518 & 1.89 & 0.688 \\
    \midrule
    EG   & 103.31\% & 0.142 & 0.421 & 0.502 & 0.386 & 0.303 \\
    \midrule
    ONS   & 93.51\% & 0.288 & -0.179 & 0.498 & -0.172 & -0.19 \\
    \midrule
    PAMR  & 116.64\% & 0.213 & 1.09 & 0.509 & 1.94 & 0.826 \\
    \midrule
    RMR   & 115.23\% & 0.146 & \underline{1.28} & 0.516 & \underline{2.00} & 1.11 \\
    \midrule
    UP & 107.17\% & 0.158 & 1.16 & 0.510 & 1.14 & 0.516 \\
    \bottomrule
    \end{tabular}}%
  \caption{Performance of CAD strategy in back-test of 800 stocks with 0.05\% commission fee.The performance metrics are Final Portfolio Value, Max Drawdown, Sharpe Ratio. The other strategies in the table are A3C-CAPM method, Buy and Hold (BAH), Uniform Constant Rebalanced portfolio (CRP) \citep{Kozat2007Universal}, Robust Median Reversion Strategy (RMR) \citep{Huang2013Robust}, Passive Aggressive Median Reversion Strategy (PAMR) \citep{Li2012PAMR}, Online Newton Selection (ONS) \citep{Agarwal2006}, Exponentiated Gradient (EG). The value which has bold font represents the best result of these strategies, the value which has underline represents the second best result of these strategies.}
  \label{tab:cl200}%
\end{table}

Table \ref{tab:cl200} presents the average back-test results based on four different time periods of four datasets, which collectively consist of 800 stocks. The metrics Final Value, Max Drawdown, Positive Days, Sharpe Ratio, Sortino Ratio, and Calmar Ratio are considered in the back-test, with a commission fee of 0.05\%. The values in bold font represent the best results among the strategies, while the underlined values represent the second-best results. The Buy and Hold strategy achieves a final value of approximately 101.54\% after 360 trading periods, indicating a relatively stable overall market trend.
The CAD strategy demonstrates the highest return (151.29\%), while the A3C-CAPM strategy, which use CAPM as hedging method, achieves the second-best return (124.06\%) in the back-test experiment. Furthermore, the CAD strategy outperforms all other strategies, including the A3C-CAPM strategy, which achieves the second-best results based on all six metrics. This superior performance of the CAD strategy can be attributed to the DBSCN cluster method, which automatically divides the stocks into different clusters, including the identification of stocks that do not belong to any cluster. As a result, the reinforcement learning algorithm can more effectively learn a suitable trading policy for stocks within the same cluster.

\section{Conclusions and Future Work}
In this paper, we propose a novel multiple-period online portfolio construction strategy. Our strategy begins by employing the t-SNE method and the DBSCAN clustering method to divide stocks into different trading pools. Subsequently, we utilize reinforcement learning with the algorithm A3C and algorithm DDPG to construct the portfolio weight vector based on these distinct clusters of stocks. Unlike other existing reinforcement learning methods that solely output the portfolio weight directly and lack flexibility, our strategy can be easily adjusted by employing different trading rules. The profitability of our strategy (151.29\%) surpasses that of all traditional portfolio management strategies (across 8 categories), as demonstrated in the paper through the average back-test results over a dataset comprising 800 different stocks in 4 distinct time periods listed in SSE and NASDAQ. The back-test results indicate that the CAD strategy outperforms the A3C-CAPM strategy, which serves as the benchmark in this research, across all six metrics. The satisfactory performance of the back-test confirms the effectiveness of the CAD strategy and suggests the potential for a new direction in portfolio management research that combines reinforcement learning with clustering methods.

The performance of the CAD strategy can be enhanced through the following approaches:
\begin{itemize}
  \item[1.] \textbf{New Indexes}: We observed that the cluster methods used in our experiments yielded better results when new indexes were incorporated. Therefore, we will expand the number of indexes employed in future experiments. Additionally, we plan to extend the time period for feature extraction and record the quarterly values of these indexes for each stock. Subsequently, we will utilize these quarterly index values in our experiments.
  \item[2.] \textbf{New Cluster methods}: In our current experiment, we compared the K-means cluster method with the DBSCAN cluster method. However, there are several other cluster methods available, such as Spectral Clustering and Hierarchical Clustering, which we intend to explore in future research.
  \item[3.] \textbf{Increase the number of stocks}: Our experiments revealed that the cluster methods failed to yield satisfactory results when applied to a small number of stocks. This may be attributed to the requirement of finding a cluster center, which becomes challenging with a limited number of stocks. Consequently, we will increase the number of stocks used in our experiments to achieve better outcomes.
\end{itemize}
\newpage
\bibliographystyle{plainnat}
\bibliography{reference_final}

\begin{thebibliography}{47}
\providecommand{\natexlab}[1]{#1}
\providecommand{\url}[1]{\texttt{#1}}
\expandafter\ifx\csname urlstyle\endcsname\relax
  \providecommand{\doi}[1]{doi: #1}\else
  \providecommand{\doi}{doi: \begingroup \urlstyle{rm}\Url}\fi

\bibitem[Agarwal et~al.(2006)Agarwal, Hazan, Kale, and Schapire]{Agarwal2006}
Amit Agarwal, Elad Hazan, Satyen Kale, and Robert~E. Schapire.
\newblock Algorithms for portfolio management based on the newton method.
\newblock In \emph{International Conference}, pages 9--16, 2006.

\bibitem[Bikker et~al.(2007)Bikker, Spierdijk, and van~der Sluis]{Jacob2007}
Jacob~A. Bikker, Laura Spierdijk, and Pieter~Jelle van~der Sluis.
\newblock Market impact costs of institutional equity trades.
\newblock \emph{Journal of International Money and Finance}, 2007.

\bibitem[Borodin et~al.(2011)Borodin, Elyaniv, and Gogan]{Borodin2011}
A.~Borodin, R.~Elyaniv, and V.~Gogan.
\newblock Can we learn to beat the best stock.
\newblock \emph{Journal of Artificial Intelligence Research}, 21\penalty0
  (1):\penalty0 579--594, 2011.

\bibitem[Boxer(2014)]{Moving2014}
H.~Boxer.
\newblock \emph{Moving Average Convergence/Divergence}.
\newblock Profitable Day and Swing Trading: Using Price/Volume Surges and
  Pattern Recognition to Catch Big Moves in the Stock Market, 2014.

\bibitem[Brockwell and Davis(2001)]{2001Discussion}
P.~J. Brockwell and R.~A. Davis.
\newblock Discussion of "non-gaussian ornstein-uhlenbeck based models and some
  of their uses in financial economics,".
\newblock 2001.

\bibitem[Cazzoli et~al.(2016)Cazzoli, Sharma, Treccani, and Lillo]{Cazzoli2016}
Lorenzo Cazzoli, Rajesh Sharma, Michele Treccani, and Fabrizio Lillo.
\newblock A large scale study to understand the relation between twitter and
  financial market.
\newblock In \emph{Network Intelligence Conference}, 2016.

\bibitem[Chen et~al.(2021)Chen, Zhang, Mehlawat, and Jia]{Chen2021Mean}
W.~Chen, H.~Zhang, M.~K. Mehlawat, and L.~Jia.
\newblock Mean variance portfolio optimization using machine learning-based
  stock price prediction.
\newblock \emph{Applied Soft Computing}, 100\penalty0 (1):\penalty0 106943,
  2021.

\bibitem[Cooper(1999)]{Cooper1999New}
Robert G.~Edgett Cooper.
\newblock New product portfolio management:.
\newblock \emph{Journal of Product Innovation Management}, 16\penalty0
  (4):\penalty0 333--351, 1999.

\bibitem[Davelos et~al.(2001)Davelos, Anita, Kinkel, Linda, Samac, and
  Deborah]{Davelos2001Modern}
Davelos, L~Anita, Kinkel, L~Linda, Samac, and A~Deborah.
\newblock \emph{Modern investment theory}.
\newblock Prentice Hall,, 2001.

\bibitem[Dempster and Leemans(2006)]{Dempster2006}
M.~A.~H. Dempster and V.~Leemans.
\newblock An automated fx trading system using adaptive reinforcement learning.
\newblock \emph{Expert Systems with Applications}, 30\penalty0 (3):\penalty0
  543--552, 2006.

\bibitem[Ester(1996{\natexlab{a}})]{1996DBSCAN}
M.~Ester.
\newblock A density-based algorithm for discovering clusters in large spatial
  databases with noise.
\newblock \emph{Proc.int.conf.knowledg Discovery \& Data Mining},
  1996{\natexlab{a}}.

\bibitem[Ester(1996{\natexlab{b}})]{Cluster1996}
M.~Ester.
\newblock A density-based algorithm for discovering clusters in large spatial
  databases with noise.
\newblock \emph{Proc.int.conf.knowledg Discovery \& Data Mining},
  1996{\natexlab{b}}.

\bibitem[Fama(1970)]{Efficient1970}
E.~Fama.
\newblock Efficient market hypothesis: A review of theory and empirical work.
\newblock 1970.

\bibitem[Foster(2015)]{A2015}
D.~J. Foster.
\newblock A model of hippocampally dependent navigation, using the temporal
  difference learning rule.
\newblock \emph{Hippocampus}, 10, 2015.

\bibitem[Funt(2010)]{Funt2010}
M.~J. Funt.
\newblock Technical analysis.
\newblock \emph{Pennsylvania Dental Journal}, 77\penalty0 (2):\penalty0 33,
  2010.

\bibitem[Gao et~al.(2020)Gao, Gao, Hu, Jiang, and Su]{2020Application}
Z.~Gao, Y.~Gao, Y.~Hu, Z.~Jiang, and J.~Su.
\newblock Application of deep q-network in portfolio management.
\newblock 2020.

\bibitem[Glabadanidis(2015)]{Market2015}
P.~Glabadanidis.
\newblock Market timing and moving averages.
\newblock 2015.

\bibitem[Gu et~al.(2016)Gu, Lillicrap, Ghahramani, Turner, and Levine]{Gu2016Q}
S.~Gu, T.~Lillicrap, Z.~Ghahramani, R.~E. Turner, and S.~Levine.
\newblock Q-prop: Sample-efficient policy gradient with an off-policy critic.
\newblock 2016.

\bibitem[Heaton et~al.(2017)Heaton, Polson, and Witte]{Heaton2017Deep}
J.~B. Heaton, N.~G. Polson, and J.~H. Witte.
\newblock Deep learning for finance: deep portfolios.
\newblock \emph{Applied Stochastic Models in Business \& Industry}, 33\penalty0
  (1), 2017.

\bibitem[Huang et~al.(2013)Huang, Zhou, Li, Hoi, and Zhou]{Huang2013Robust}
Dingjiang Huang, Junlong Zhou, Bin Li, Steven C.~H. Hoi, and Shuigeng Zhou.
\newblock Robust median reversion strategy for on-line portfolio selection.
\newblock In \emph{International Joint Conference on Artificial Intelligence},
  volume~28, pages 2006--2012, 2013.

\bibitem[Jarrow and Turnbull(2012)]{2012Pricing}
R.~A. Jarrow and S.~M. Turnbull.
\newblock Pricing derivatives on financial securities subject to credit risk.
\newblock \emph{Journal of Finance}, 50\penalty0 (1):\penalty0 53--85, 2012.

\bibitem[Jensen et~al.(1973)Jensen, Black, and Scholes]{1973The}
M.~C. Jensen, F.~Black, and M.~S. Scholes.
\newblock The capital asset pricing model: Some empirical tests.
\newblock \emph{Social Science Electronic Publishing}, 1973.

\bibitem[Jeremy et~al.(2014)Jeremy, Serror, Denis, S., and
  Grebenkov]{Jeremy2014}
Jeremy, Serror, Denis, S., and Grebenkov.
\newblock Following a trend with an exponential moving average: Analytical
  results for a gaussian model.
\newblock \emph{Physica, A. Statistical mechanics and its applications},
  394:\penalty0 288--303, 2014.

\bibitem[Jiang and Coenen(2019)]{Jiang2018Long}
Zhengyong Jiang and Frans Coenen.
\newblock Long short-term memory-based multi-period price prediction for
  portfolio management.
\newblock In Petra Perner, editor, \emph{Machine Learning and Data Mining in
  Pattern Recognition, 15th International Conference on Machine Learning and
  Data Mining, {MLDM} 2019, New York, NY, USA, July 20-25, 2019, Proceedings,
  Volume {I}}, pages 187--200. ibai publishing, 2019.

\bibitem[Kaiser et~al.(2019{\natexlab{a}})Kaiser, Babaeizadeh, Milos, Osinski,
  Campbell, Czechowski, Erhan, Finn, Kozakowski, and Levine]{Kaiser2019Model}
L.~Kaiser, M.~Babaeizadeh, P.~Milos, B.~Osinski, R.~H. Campbell, K.~Czechowski,
  D.~Erhan, C.~Finn, P.~Kozakowski, and S.~Levine.
\newblock Model-based reinforcement learning for atari.
\newblock 2019{\natexlab{a}}.

\bibitem[Kaiser et~al.(2019{\natexlab{b}})Kaiser, Babaeizadeh, Milos, Osinski,
  Campbell, Czechowski, Erhan, Finn, Kozakowski, and Levine]{Model2019}
L.~Kaiser, M.~Babaeizadeh, P.~Milos, B.~Osinski, R.~H. Campbell, K.~Czechowski,
  D.~Erhan, C.~Finn, P.~Kozakowski, and S.~Levine.
\newblock Model-based reinforcement learning for atari.
\newblock 2019{\natexlab{b}}.

\bibitem[Kozat and Singer(2007)]{Kozat2007Universal}
Suleyman~S. Kozat and Andrew~C. Singer.
\newblock Universal constant rebalanced portfolios with switching.
\newblock In \emph{IEEE International Conference on Acoustics, Speech and
  Signal Processing}, pages III--1129 -- III--1132, 2007.

\bibitem[Li and Hoi(2012{\natexlab{a}})]{Li20125}
Bin Li and Steven C.~H. Hoi.
\newblock Online portfolio selection: A survey.
\newblock \emph{Papers}, 46\penalty0 (3):\penalty0 1--36, 2012{\natexlab{a}}.

\bibitem[Li and Hoi(2012{\natexlab{b}})]{Li2012On}
Bin Li and Steven C.~H. Hoi.
\newblock On-line portfolio selection with moving average reversion.
\newblock \emph{Papers}, pages 173\--{}190, 2012{\natexlab{b}}.

\bibitem[Li and Hoi(2012{\natexlab{c}})]{Li2012Online}
Bin Li and Steven C.~H. Hoi.
\newblock Online portfolio selection: A survey.
\newblock \emph{Acm Computing Surveys}, 46\penalty0 (3):\penalty0 1--36,
  2012{\natexlab{c}}.

\bibitem[Li et~al.(2012)Li, Zhao, Hoi, and Gopalkrishnan]{Li2012PAMR}
Bin Li, Peilin Zhao, Steven~C. Hoi, and Vivekanand Gopalkrishnan.
\newblock Pamr: Passive aggressive mean reversion strategy for portfolio
  selection.
\newblock \emph{Machine Learning}, 87\penalty0 (2):\penalty0 221--258, 2012.

\bibitem[Li(2016)]{Li2016}
Zhiyong Li.
\newblock Maxdrawdown: Stata module to calculate the maximum drawdown of a
  stock, fund or other financial product.
\newblock \emph{Statistical Software Components}, 2016.

\bibitem[Lillicrap et~al.(2015)Lillicrap, Hunt, Pritzel, Heess, Erez, Tassa,
  Silver, and Wierstra]{2015Continuous}
Timothy~P. Lillicrap, Jonathan~J. Hunt, Alexander Pritzel, Nicolas Heess, Tom
  Erez, Yuval Tassa, David Silver, and Daan Wierstra.
\newblock Continuous control with deep reinforcement learning.
\newblock \emph{Computer ence}, 2015.

\bibitem[Markowitz(1952)]{Markowitz1952Portfolio}
Harry Markowitz.
\newblock Portfolio selection.
\newblock \emph{Journal of Finance}, 7\penalty0 (1):\penalty0 77--91, 1952.

\bibitem[Memmel(2003)]{Memmel2003}
Christoph Memmel.
\newblock Performance hypothesis testing with the sharpe ratio.
\newblock \emph{Social Science Electronic Publishing}, 27\penalty0
  (3):\penalty0 299--306, 2003.

\bibitem[Mnih et~al.(2015)Mnih, Kavukcuoglu, Silver, Rusu, Veness, Bellemare,
  Graves, Riedmiller, Fidjeland, and Ostrovski]{Mnih2015}
V~Mnih, K~Kavukcuoglu, D~Silver, A.~A. Rusu, J~Veness, M.~G. Bellemare,
  A~Graves, M~Riedmiller, A.~K. Fidjeland, and G~Ostrovski.
\newblock Human-level control through deep reinforcement learning.
\newblock \emph{Nature}, 518\penalty0 (7540):\penalty0 529, 2015.

\bibitem[Moody and Saffell(2001)]{Moody2001}
J~Moody and M~Saffell.
\newblock Learning to trade via direct reinforcement.
\newblock \emph{Neural Networks IEEE Transactions on}, 12\penalty0
  (4):\penalty0 875--889, 2001.

\bibitem[Murto and Tervioe(2014)]{Exit2014}
P.~Murto and Marko Tervioe.
\newblock Exit options and dividend policy under liquidity constraints.
\newblock \emph{International Economic Review}, 55\penalty0 (1):\penalty0
  197--221, 2014.

\bibitem[Park and Hastie(2007)]{L12007}
M.~Y. Park and T.~Hastie.
\newblock L1-regularization path algorithm for generalized linear models.
\newblock \emph{Journal of the Royal Statistical Society: Series B (Statistical
  Methodology)}, 2007.

\bibitem[Sharpe(1964)]{William1964}
William~F. Sharpe.
\newblock Capital asset prices: A theory of market equilibrium under conditions
  of risk.
\newblock \emph{The Journal of Finance}, 19\penalty0 (3):\penalty0 425--442,
  1964.

\bibitem[Sharpe(1994)]{Sharpe1994}
William~F. Sharpe.
\newblock The sharpe ratio.
\newblock \emph{Journal of Portfolio Management}, 21\penalty0 (1):\penalty0
  49--58, 1994.

\bibitem[Sutton et~al.(1999)Sutton, Mcallester, Singh, and Mansour]{Policy1999}
Richard~S. Sutton, David Mcallester, Satinder Singh, and Yishay Mansour.
\newblock Policy gradient methods for reinforcement learning with function
  approximation.
\newblock \emph{Submitted to Advances in Neural Information Processing
  Systems}, 12, 1999.

\bibitem[Tadepalli and Ok(2007)]{Tadepalli2007Model}
P.~Tadepalli and D.~K. Ok.
\newblock Model-based reinforcement learning.
\newblock \emph{Artificial Intelligence}, 100\penalty0 (1-2):\penalty0
  177--224, 2007.

\bibitem[Van~der Maaten~L(2008)]{2008Visualizing}
Hinton~G Van~der Maaten~L.
\newblock Visualizing data using t-sne[j].
\newblock \emph{Journal of Machine Learning Research}, 9:\penalty0 2579--2605,
  2008.

\bibitem[Wei et~al.(2015)Wei, Liu, and Shi]{A20152}
Q.~Wei, D.~Liu, and G.~Shi.
\newblock A novel dual iterative $q$-learning method for optimal battery
  management in smart residential environments.
\newblock \emph{IEEE Transactions on Industrial Electronics}, 62\penalty0
  (4):\penalty0 2509--2518, 2015.

\bibitem[Xiong et~al.(2016)Xiong, Nichols, and Shen]{Xiong2016Deep}
Ruoxuan Xiong, Eric~P. Nichols, and Yuan Shen.
\newblock Deep learning stock volatility with google domestic trends.
\newblock \emph{Papers}, 2016.

\bibitem[Zou(2018)]{Predicting2018}
D.~W. Zou.
\newblock Predicting stock price movements from annual reports.
\newblock 2018.

\end{thebibliography}

\section{Appendix}
The appendix shows the detailed stock code or logogram of stocks which are used in the back-test.
\begin{itemize}
    \item[1.] \textbf{SSE}. There are 400 stocks from SSE, their stock codes are 600000, 600004, 600015, 600016, 600019, 600021, 600028, 600031, 600036, 600037, 600048, 600050, 600058, 600059, 600060, 600073, 600089, 600100, 600101, 600102, 600103, 600104, 600105, 600106, 600107, 600108, 600109, 600110, 600111, 600112, 600113, 600114, 600115, 600116, 600117, 600118, 600119, 600120, 600121, 600122, 600123, 600143, 600153, 600156, 600158, 600159, 600160, 600161, 600162, 600163, 600165, 600166, 600167, 600168, 600169, 600170, 600171, 600172, 600173, 600175, 600176, 600177, 600178, 600179, 600180, 600182, 600183, 600184, 600185, 600186, 600187, 600188, 600189, 600190, 600191, 600192, 600193, 600195, 600196, 600197, 600198, 600199, 600200, 600201, 600202, 600203, 600206, 600207, 600208, 600209, 600210, 600211, 600212, 600213, 600215, 600216, 600217, 600218, 600219, 600220, 600221, 600222, 600223, 600226, 600227, 600228, 600229, 600230, 600231, 600232, 600233, 600234, 600235, 600236, 600237, 600238, 600239, 600240, 600241, 600243, 600246, 600247, 600248, 600249, 600250, 600251, 600252, 600255, 600256, 600257, 600258, 600259, 600260, 600261, 600262, 600265, 600266, 600267, 600268, 600269, 600270, 600271, 600272, 600273, 600275, 600276, 600277, 600278, 600279, 600280, 600281, 600282, 600283, 600284, 600285, 600287, 600288, 600289, 600290, 600291, 600292, 600293, 600295, 600297, 600298, 600299, 600300, 600301, 600302, 600303, 600305, 600306, 600307, 600308, 600309, 600310, 600311, 600312, 600313, 600315, 600316, 600317, 600318, 600319, 600320, 600321, 600322, 600323, 600325, 600326, 600327, 600328, 600329, 600330, 600331, 600332, 600333, 600335, 600336, 600337, 600338, 600339, 600340, 600343, 600345, 600346, 600348, 600350, 600351, 600352, 600353, 600354, 600355, 600356, 600358, 600359, 600360, 600361, 600362, 600363, 600365, 600366, 600367, 600368, 600369, 600370, 600371, 600372, 600373, 600375, 600376, 600377, 600378, 600379, 600380, 600381, 600382, 600383, 600385, 600386, 600387, 600388, 600389, 600390, 600391, 600392, 600393, 600395, 600396, 600397, 600400, 600403, 600405, 600406, 600408, 600409, 600410, 600415, 600416, 600418, 600419, 600420, 600422, 600423, 600425, 600426, 600428, 600429, 600433, 600435, 600436, 600438, 600439, 600444, 600446, 600448, 600449, 600452, 600456, 600458, 600459, 600460, 600461, 600463, 600466, 600467, 600468, 600469, 600470, 600475, 600476, 600477, 600478, 600479, 600480, 600481, 600482, 600483, 600486, 600487, 600488, 600489, 600491, 600493, 600495, 600496, 600497, 600498, 600499, 600500, 600501, 600502, 600503, 600505, 600506, 600507, 600508, 600509, 600510, 600511, 600512, 600513, 600516, 600517, 600518, 600519, 600520, 600521, 600522, 600523, 600525, 600526, 600527, 600528, 600529, 600530, 600531, 600532, 600533, 600535, 600536, 600537, 600538, 600539, 600540, 600543, 600545, 600546, 600547, 600549, 600550, 600551, 600552, 600555, 600557, 600558, 600559, 600560, 600561, 600562, 600563, 600565, 600566, 600567, 600568, 600569, 600570, 600571, 600572, 600573, 600575, 600576, 600577, 600578, 600579, 600580, 600581, 600582, 600583, 600584, 600585, 600586, 600587, 600588, 600589, 600590, 600592, 600593, 600594, 600595, 600596, 600597, 600598, 600599, 600600, 600601, 600602, 600603, 600604, 600605.
    \item[2.] \textbf{NASDAQ}. There are 400 stocks from NASDAQ, their logograms are A, AA, AAPL, ABC, ABT, ACN, ADBE, ADI, ADM, ADP, ADSK, AEE, AEP, AES, AFL, AGN, AIG, AIV, AIZ, AKAM, AKS, ALL, AMAT, AMD, AMGN, AMP, AMT, AMZN, AN, ANF, AON, APA, APD, APH, ATI, AVB, AVP, AVY, AXP, AZO, BA, BAC, BAX, BBBY, BBT, BBY, BDX, BEN, BIG, BIIB, BK, BLK, BLL, BMY, BSX, BX, BXP, C, CACI, CAG, CAH, CAR, CAT, CB, CCE, CCL, CELG, CERN, CF, CHK, CHRW, CI, CIF, CINF, CL, CLAR, CLF, CLX, CMA, CMCSA, CME, CMG, CMI, CMS, CNP, CNX, COG, COP, COST, CPB, CRM, CSCO, CSX, CTAS, CTL, CTSH, CTXS, CVCO, CVS, CVX, D, DD, DE, DF, DFS, DGX, DHI, DHR, DIS, DISCA, DLR, DNR, DO, DOV, DRAD, DRI, DTE, DTF, DTH, DUK, DVA, DVN, EA, EBAY, ECL, ED, EEB, EFX, EIX, EL, EMN, EMR, EMX, EOG, EOS, EQR, EQT, ERIC, ETFC, ETN, ETR, EW, EWA, EWC, EXC, EXPD, EXPE, F, FAST, FCX, FDL, FDX, FE, FFIV, FHN, FII, FIS, FISV, FITB, FLIR, FLR, FLS, FMC, FPX, FRXX, FSLR, FTI, FTR, GD, GE, GILD, GIS, GLW, GME, GNW, GOOG, GPC, GPN, GPS, GS, GT, GWW, HAL, HAS, HBAN, HD, HES, HIG, HOG, HON, HP, HPI, HPQ, HR, HRB, HRL, HST, HSY, HUM, IBM, ICE, IFF, IGD, IGT, INTC, INTU, IP, IPG, IR, IRM, ISRG, ITT, ITW, IVZ, JBL, JBSS, JCI, JCP, JEC, JNJ, JNPR, JOBS, JPM, JWN, K, KEY, KIM, KLAC, KMB, KMX, KO, KR, KSS, L, LAD, LEG, LEN, LH, LLY, LM, LMT, LNC, LOW, LSI, LUV, M, MA, MAR, MAS, MAT, MCD, MCHP, MCK, MCO, MDT, MDU, MET, MKC, MMC, MMM, MO, MRK, MRO, MS, MSFT, MSI, MTB, MU, MUR, MYL, NBL, NBR, NDAQ, NE, NEE, NEM, NFLX, NG, NHC, NI, NKE, NOC, NOV, NRG, NSC, NTAP, NTRS, NUE, NVDA, NWL, OI, OKE, OMC, ORCL, ORLY, PAYX, PBCT, PBI, PCAR, PCG, PDCO, PEG, PEP, PFE, PFG, PG, PGR, PKI, PLD, PNC, PNW, PPG, PPL, PPT, PRU, PSA, PWR, PXD, QCOM, R, RF, RHI, RL, ROK, ROP, ROST, RRC, RRD, RSG, RTN, S, SAN, SBUX, SCHW, SEE, SHW, SJM, SLB, SLM, SNA, SO, SPG, SRCL, SRE, STI, STT, STZ, SUN, SWK, SWN, SYK, T, TAP, TDC, TER, TGT, THC, TIF, TJX, TMO, TROW, TRV, TSN, TXN, TXT, UNH, UNM, UNP, UPS, URBN, USB, UTX, VAR, VFC, VIA, VLO, VMC, VNO, VRSN, VTR, VZ, WAT, WDC, WEC, WFC, WHR, WM, WMB, WMT, WU, WY, WYNN, X, XEL, XLNX, XOM, XRAY, XRX, YUM, ZION.
\end{itemize}

\end{document}